\documentclass{article}
\usepackage{amsfonts}
\usepackage{makeidx}
\usepackage{amssymb}
\usepackage{amsmath}
\usepackage{amstext}
\usepackage{graphicx}
\usepackage{latexsym}
\usepackage{amsthm}

\newtheorem{theorem}{Theorem}

\newtheorem{corollary}[theorem]{Corollary}

\newtheorem{definition}[theorem]{Definition}

\newtheorem{lemma}[theorem]{Lemma}

\newtheorem{proposition}[theorem]{Proposition}
\newtheorem{remark}[theorem]{Remark}

\input tcilatex
\begin{document}

\title{A classical model for the Maxwell equations coupled with matter\\
}
\author{Vieri Benci$^{\ast }$, \\
$^{\ast}$Dipartimento di Matematica Applicata \textquotedblleft U.
Dini\textquotedblright\\
Universit\`{a} degli Studi di Pisa \\
Via Filippo Buonarroti 1/c, 56127 Pisa, Italy\\
e-mail: benci@dma.unipi.it \\
\medskip \\
\textit{To the memory of my friend Antonio Ambrosetti}}
\maketitle

\begin{abstract}
We present a simple model of interaction of the Maxwell equations with a 
\textit{matter field} defined by the Klein-Gordon equation. A simple linear
interaction and a nonlinear perurbation produce solutions of the equations
containing hylomorphic solitons, namely stable, solitary waves whose
existence is related to the ratio energy/charge. These solitons, at low
energy, behave as poinwise charged particles in an electromagnetic field.

\medskip

\textbf{Key words:} Maxwell equations, Nonlinear Klein-Gordon equation,
solitons, Q-balls, variational methods.

\ 

\textbf{AMS Subject Classification}: 35C08, 35A15, 37K40, 78M30.
\end{abstract}

\tableofcontents

\section{Introduction}

In classical mechanics the coupling of the electromagnetic field is given by
the following equation:%
\begin{equation}
\frac{d}{dt}\left( m\dot{\xi}\right) =q\left( \mathbf{E}+\dot{\xi}\times 
\mathbf{H}\right)  \label{LL}
\end{equation}%
where $m$ is the mass of a particle, $\xi =\xi (t)$ is its position in space
and $\dot{\xi}$ is the time derivative of $\xi $. Unfortunately this
equation is not consistent with the Maxwell equations. One of the main
reasons of this inconsistency comes from the fact that the Maxwell equations
are relativistic invariant and hence the inertial mass/energy of a charged
material point is infinite. If the material point is replaced by a sort of
ball other problems are present such as the selfinteraction of the field
produced by the particle and the difficulty of a relativistic description of
a solid body. As far as I know there is not a satisfactory description of
the dynamics of a microscopic charged "ball" in an e.m. field. With the
advent of quantum mechanics, this problem has lost its relevance and quantum
models have been sought to describe this interaction.

Here we recall the formally simplest of them, since it has some relevance
for this paper. It is given by the interaction of the Klein-Gordon equation
(which describes a spinless boson field) with the e.m. field. In this case
the action functional is given by%
\begin{equation}
\mathbb{A}_{\text{\textsc{w}}}:=\frac{1}{2}\iint \left( \left\vert \left(
\partial _{t}+iq\varphi \right) \psi \right\vert ^{2}-\left\vert \left(
\nabla -iq\mathbf{A}\right) \psi \right\vert ^{2}+m^{2}\left\vert \psi
\right\vert ^{2}\right) dx\ dt.  \label{W}
\end{equation}%
where $\psi $ is the wave-function of the boson field, $(\varphi ,\mathbf{A}%
) $ is the gauge potential, $m$ and $q$ is the mass and the electric charge
of a particle.

Despite the fact that quantum electrodynamics (QED) is a well-established
theory, we think that the study of the possibility of a consistent classical
electrodynamics (CED) is still a relevant issue that might shed new light
also on the unsolved problems in QED.

The model proposed here is based on the idea that charged particles can be
described by solitons which can be seen as bumps of a "matter field". The
idea is not new; in the last half century, since the pioneering work of
Rosen in 68 \cite{rosen68}, a lot of papers have been written. The original
point of this paper is the introduction of a very simple interaction between
the "matter field" and the e.m. field which produces identical particles
which obey the known laws of CED.

\section{The model\label{mod}}

\subsection{The basic equations}

The action of the electromagnetic field is defined by the Lagrangian density 
\begin{equation}
\mathcal{L}_{\text{\textsc{f}}}\left[ \varphi ,\mathbf{A}\right] =\frac{1}{2}%
\left( \left\vert \partial _{t}\mathbf{A}+\nabla \varphi \right\vert
^{2}-\left\vert \nabla \times \mathbf{A}\right\vert ^{2}\right)  \label{1}
\end{equation}%
Now, we need to choose an equation to describe the matter field. The
simplest semilinear equation invariant for the Poincar\'{e} group is the
following 
\begin{equation}
\square \psi +W^{\prime }(\left\vert \psi \right\vert )\frac{\psi }{%
\left\vert \psi \right\vert }=0  \label{NKG}
\end{equation}%
where 
\begin{equation*}
\square :=\partial _{t}^{2}-\Delta ;
\end{equation*}%
$\psi $ takes values in $\mathbb{C}$ and 
\begin{equation}
W(s)=\frac{1}{2}s^{2}+N(s);  \label{WN}
\end{equation}%
\begin{equation*}
N\in C^{2}\left( \mathbb{R}^{3}\right) ,\ \ N(0)=N^{\prime }(0)=0.
\end{equation*}%
The use of the complex variable is important since it gives to the field $%
\psi $ an internal degree of freedom represented by a phase shift given by 
\begin{equation}
\psi (t,x)\mapsto e^{i\theta }\psi (t,x)  \label{trg}
\end{equation}%
Usually, people refer to equation (\ref{NKG}) as to the nonlinear
Klein-Gordon equation since its linearization gives the Klein-Gordon
equation:%
\begin{equation}
\square \psi +\psi =0  \tag{KG}
\end{equation}%
Equation (\ref{NKG}) has a variational structure and its Lagrangian density
can be written as follows:%
\begin{eqnarray}
\mathcal{L}_{\text{\textsc{m}}}\left[ u,S\right] &=&\frac{1}{2}\left(
\left\vert \partial _{t}\psi \right\vert ^{2}-\left\vert \nabla \psi
\right\vert ^{2}\right) -W(\left\vert \psi \right\vert )  \notag \\
&=&\frac{1}{2}\left[ \left\vert \partial _{t}u\right\vert ^{2}-\left\vert
\nabla u\right\vert ^{2}+\left( \partial _{t}S\right) ^{2}u^{2}-\left\vert
\nabla S\right\vert ^{2}u^{2}\right] -W(u)  \label{2}
\end{eqnarray}%
where we have set%
\begin{equation}
\psi (t,x)=u(t,x)e^{iS(t,x)};\ \ u\geq 0.  \label{3}
\end{equation}

We want to couple the matter field $\psi $ with the electromagnetic field in
the most simple and natural way, taking care that the Lorentz invariance of
the equations be satisfied. Since the Lagrangian of the electromagnetic
field depend on the $4$-vector $(\varphi ,\mathbf{A)}$, it must be coupled
with a $4$-vector determined by $\psi $. There are two possible candidates
which are linear in $u$ and invariant for the transformation (\ref{3}):%
\begin{equation*}
\left( \partial _{t}u,\nabla u\right) 
\end{equation*}%
\begin{equation*}
\left( \partial _{t}S,\nabla S\right) u
\end{equation*}%
Notice that these vectors are controvariant. They lead to the following
interaction Lagrangian densities:%
\begin{equation*}
\mathcal{L}_{0}\left[ u,\varphi ,\mathbf{A}\right] =\beta \left( \varphi
\partial _{t}u+\mathbf{\QTR{mathbf}{A}}\cdot \nabla u\right) 
\end{equation*}%
\begin{equation}
\mathcal{L}_{\text{\textsc{i}}}\left[ u,S,\varphi ,\mathbf{A}\right] =\beta
\left( \partial _{t}S\varphi +\mathbf{\QTR{mathbf}{A}}\cdot \nabla S\right) u
\label{Li}
\end{equation}%
where $\beta $ is the interaction constant which, in order to fix the ideas,
we assume positive; on the contrary, the sign "$+$" in the above definitions
is necessary since $\left( \partial _{t}u,\nabla u\right) $ and $\left(
\partial _{t}S,\nabla S\right) $ are covariant with respect to the
time-coordinate; a "$-$" would violate the time-reversal property and the
equations would lose the invariance for the Poincar\'{e} group. Since $%
\mathcal{L}_{\text{\textsc{i}}}$ is not locally gauge invariant, we assume
the Lorentz condition%
\begin{equation}
\partial _{t}\varphi +\nabla \cdot \mathbf{A}=0.  \label{LC}
\end{equation}%
In this paper we will examine the case $\mathcal{L}_{\text{\textsc{i}}}$
(with $\beta >0)$ which provides a very rich model.

So, we will study the equations relative to the following Lagrangian density:%
\begin{eqnarray*}
\mathcal{L} &=&\mathcal{L}_{\text{\textsc{m}}}+\mathcal{L}_{\text{\textsc{i}}%
}+\mathcal{L}_{\text{\textsc{f}}} \\
&=&\frac{1}{2}\left[ \left\vert \partial _{t}u\right\vert ^{2}-\left\vert
\nabla u\right\vert ^{2}\right] dxdt-W(u)+\frac{1}{2}\left[ \left( \partial
_{t}S\right) ^{2}-\left\vert \nabla S\right\vert ^{2}\right] u^{2} \\
&&+\beta \left( \mathbf{\QTR{mathbf}{A}}\cdot \nabla S+\varphi \partial
_{t}S\right) u \\
&&+\frac{1}{2}\left( \left\vert \partial _{t}\mathbf{A}+\nabla \varphi
\right\vert ^{2}-\left\vert \nabla \times \mathbf{A}\right\vert ^{2}\right)
\end{eqnarray*}%
Making the variation of the action functional 
\begin{equation*}
\mathbb{A}=\iint \mathcal{L\ }dx\ dt
\end{equation*}%
with respect to $u,\ S,$ $\varphi $ and $\mathbf{A}$ we get the following
system of equations:%
\begin{equation}
\square u+W^{\prime }(u)-\left[ \left( \partial _{t}S\right) ^{2}-\left\vert
\nabla S\right\vert ^{2}\right] u=\beta \left( \mathbf{\QTR{mathbf}{A}}\cdot
\nabla S+\varphi \partial _{t}S\right)  \label{e1}
\end{equation}%
\begin{equation}
\partial _{t}\left( \partial _{t}S\ u^{2}-\beta \varphi u\right) -\nabla
\cdot \left( \nabla S\ u^{2}+\beta \mathbf{\QTR{mathbf}{A}}u\right) =0
\label{e2}
\end{equation}%
\begin{equation}
\nabla \cdot \left( \partial _{t}\mathbf{\QTR{mathbf}{A}}+\nabla \varphi
\right) =\beta \partial _{t}Su  \label{e3}
\end{equation}%
\begin{equation}
\nabla \times \left( \nabla \times \mathbf{A}\right) +\partial _{t}\left(
\partial _{t}\mathbf{A}+\nabla \varphi \right) =\beta \nabla Su.  \label{e4}
\end{equation}%
We can express these equations with new variables in order to make the
equations independent of $\beta $ and to get the Maxwell equations: 
\begin{equation}
\mathbf{E=-}\partial _{t}\mathbf{A}-\nabla \varphi  \label{pos1}
\end{equation}%
\begin{equation}
\mathbf{H}=\nabla \times \mathbf{A}  \label{pos2}
\end{equation}%
\begin{equation}
\rho =-\beta \partial _{t}Su  \label{ecar}
\end{equation}%
\begin{equation}
\mathbf{j}=\beta \nabla Su  \label{eflu}
\end{equation}%
So we get: 
\begin{equation}
\nabla \cdot \mathbf{E}=\rho  \tag{\QTR{sc}{gauss}}  \label{gauss}
\end{equation}%
\begin{equation}
\nabla \times \mathbf{H}-\partial _{t}\mathbf{E}=\mathbf{j} 
\tag{\QTR{sc}{ampere}}  \label{amp}
\end{equation}%
and (\ref{pos1}) and (\ref{pos2}) give rise to the first couple of the
Maxwell equations: 
\begin{equation}
\nabla \times \mathbf{E}+\partial _{t}\mathbf{H}=0  \tag{\QTR{sc}{faraday}}
\end{equation}%
\begin{equation}
\nabla \cdot \mathbf{H}=0.  \tag{\QTR{sc}{gauss for magnetism}}
\end{equation}%
The equations of the matter field become:%
\begin{equation}
\square u+W^{\prime }(u)+\frac{\mathbf{j}^{2}-\rho ^{2}+\varphi \rho +%
\mathbf{\QTR{mathbf}{A}}\cdot \mathbf{j}}{u}=0  \label{brutta}
\end{equation}%
\begin{equation}
\partial _{t}\left( \rho u-\varphi u\right) +\nabla \cdot \left( \mathbf{j}u+%
\mathbf{\QTR{mathbf}{A}}u\right) =0.  \label{bruttina}
\end{equation}%
Notice that also the equations (\ref{brutta}) and (\ref{bruttina}) depend
only from the gauge independent variables $\left( u,\rho ,\mathbf{j},\mathbf{%
E},\mathbf{H}\right) $ since the dependence from $\varphi $ and $\mathbf{A}$
in these equations can be eliminated via the \textsc{gauss}$\mathfrak{\ }$%
equations which gives $\varphi $ and $\mathbf{A}$ by the appropriate Green
functions. The action can be rewritten as follows:%
\begin{eqnarray}
\mathbb{A} &=&\frac{1}{2}\iint \left( \left\vert \partial _{t}u\right\vert
^{2}-\left\vert \nabla u\right\vert ^{2}\right) dxdt-\iint W(u)dxdt  \notag
\\
&&+\frac{1}{2}\iint \left( \rho ^{2}-\mathbf{j}^{2}\right) \ dxdt  \notag \\
&&+\iint \left( \mathbf{\QTR{mathbf}{A}}\cdot \mathbf{j}-\varphi \rho
\right) \ dxdt+\frac{1}{2}\iint \left( \mathbf{E}^{2}-\mathbf{H}^{2}\right)
dxdt.  \label{bella}
\end{eqnarray}

\begin{remark}
As we have already remarked, the Lagrangian (\ref{Li}) is not invariant for
the local action of the gauge group of the electromagnetic field; however,
it is invariant for the global action of a gauge tranformation namely
invariant for the $4$-paremeters group 
\begin{equation}
(u,S,\varphi ,\mathbf{\QTR{mathbf}{A}})\mapsto \left( u,S-at-\mathbf{b}\cdot
x,\ \varphi +a,\ \mathbf{\QTR{mathbf}{A}}+\mathbf{b}\right) ;\ (a,\mathbf{b}%
)\in \mathbb{R}^{4}  \label{gau}
\end{equation}%
Moreover, we have the invariance (\ref{trg}) which can be rewritten as
follows: 
\begin{equation}
T_{\theta }S=S+\theta ;\ \ \theta \in \frac{\mathbb{R}}{2\pi \mathbb{Z}}
\label{hy}
\end{equation}%
and it is tipical of the (\ref{NKG}) equation. Notice that in this model the
invariance (\ref{hy}) is idependent of the gauge invariance (\ref{gau}) and
hence it leads to a different consrvetion law (see Section \ref{im}).
Finally, we remark that the position (\ref{ecar}) and (\ref{eflu}) are
appropriate if we put ourselves in a gauge where $(u,S,\varphi ,\mathbf{%
\QTR{mathbf}{A}})$ vanish at infinity, or to be more precise, if $u,S\in
H^{1}(\mathbb{R}^{3})$, $\varphi \in \mathcal{D}^{1,2}(\mathbb{R}^{3})$ and $%
\mathbf{\QTR{mathbf}{A}}\in \mathcal{D}^{1,2}(\mathbb{R}^{3},\mathbb{R}^{3})$
(see (\ref{D})).
\end{remark}

\subsection{Conservation laws\label{im}}

Let us examine the main integral of motion that will be used in the
following. We assume that $(u,S,\varphi ,\mathbf{A})$ is a solution of (\ref%
{e1}),...,(\ref{e4}) and that all the quantities $\left\vert \partial
_{t}u\right\vert ^{2},\ \left\vert \nabla u\right\vert ^{2},\ \left(
\partial _{t}S\right) ^{2},\ $etc. be integrable; under these assumptions we
will compute some integral of motion relevant for this paper.

\bigskip

\textbf{Conservation of Energy. } Energy, by definition, is the quantity
which is preserved by the time invariance of the Lagrangian. We have the
following result:

\begin{proposition}
\label{E}The energy takes the following form%
\begin{equation}
E=E_{\text{\textsc{m}}}+E_{\text{\textsc{f}}}+E_{\text{\textsc{i}}}
\label{sp11}
\end{equation}%
where 
\begin{eqnarray}
E_{\text{\textsc{m}}}\left[ u,S\right] &=&\frac{1}{2}\int \left[ \left\vert
\partial _{t}u\right\vert ^{2}+\left\vert \nabla u\right\vert ^{2}\right] dx+%
\frac{1}{2}\int \left[ \left( \partial _{t}S\right) ^{2}+\left\vert \nabla
S\right\vert ^{2}\right] u^{2}dx+\int W(u)dx  \notag \\
&=&\frac{1}{2}\int \left[ \left\vert \partial _{t}u\right\vert
^{2}+\left\vert \nabla u\right\vert ^{2}\right] dx+\frac{1}{2}\int \frac{%
\rho ^{2}+\mathbf{\mathbf{j}}^{2}}{\beta ^{2}}\ dx+\int W(u)dx  \label{me}
\end{eqnarray}%
is the \textbf{matter field energy},%
\begin{eqnarray*}
E_{\text{\textsc{f}}}\left[ \varphi ,\mathbf{A}\right] &=&\frac{1}{2}\int
\left( \left\vert \partial _{t}\mathbf{A}+\nabla \varphi \right\vert
^{2}+\left\vert \nabla \times \mathbf{A}\right\vert ^{2}\right) dx \\
&=&\frac{1}{2}\int \left( \mathbf{E}^{2}+\mathbf{H}^{2}\right) dx.
\end{eqnarray*}%
is the \textbf{e.m. field energy}, and%
\begin{eqnarray}
E_{\text{\textsc{i}}}\left[ u,S,\varphi ,\mathbf{A}\right] &=&-\beta \int
\left( \varphi \partial _{t}S+\mathbf{\mathbf{\QTR{mathbf}{A}}}\cdot \nabla
S\right) udx  \label{te} \\
&=&\int \left( \varphi \rho -\mathbf{\mathbf{\QTR{mathbf}{A}}\cdot \mathbf{j}%
}\right) \ dx.  \notag
\end{eqnarray}%
is the \textbf{interaction energy}.
\end{proposition}

The names \textit{matter field energy, e.m. field energy, }and\textit{\
interaction energy} are motivated by the fact that $E_{\text{\textsc{m}}}$
depend only on the matter variables $\left( u,S\right) ,$ $E_{\text{\textsc{f%
}}}$ depend on the e.m. field variables $\left( \varphi ,\mathbf{A}\right) $
and only $E_{\text{\textsc{i}}}$ depend on all the four variables.

\textbf{Proof}: By Noether's Theorem, we have that the energy density $E_{%
\text{\textsc{m}}}$ relative to the Lagrangian density $\mathcal{L}_{\text{%
\textsc{m}}}$ is given by

\begin{eqnarray*}
E_{\text{\textsc{m}}} &=&\frac{\partial \mathcal{L}_{\text{\textsc{m}}}}{%
\partial \left( \partial _{t}u\right) }\cdot \partial _{t}u+\frac{\partial 
\mathcal{L}_{\text{\textsc{m}}}}{\partial \left( \partial _{t}S\right) }%
\cdot \partial _{t}S-\mathcal{L}_{\text{\textsc{m}}} \\
&=&\left( \partial _{t}u\right) ^{2}+\left( \partial _{t}S\right) ^{2}u^{2}-%
\frac{1}{2}\left[ \left\vert \partial _{t}u\right\vert ^{2}-\left\vert
\nabla u\right\vert ^{2}-W(u)+\left( \partial _{t}S\right)
^{2}u^{2}-\left\vert \nabla S\right\vert ^{2}u^{2}\right] \\
&=&\frac{1}{2}\left[ \left\vert \partial _{t}u\right\vert ^{2}+\left\vert
\nabla u\right\vert ^{2}+\left( \partial _{t}S\right) ^{2}u^{2}+\left\vert
\nabla S\right\vert ^{2}u^{2}\right] +W(u) \\
&=&\frac{1}{2}\left[ \left\vert \partial _{t}u\right\vert ^{2}+\left\vert
\nabla u\right\vert ^{2}+\frac{\rho ^{2}+\mathbf{\mathbf{j}}^{2}}{\beta ^{2}}%
\right] +W(u)
\end{eqnarray*}%
The computation of energy density relative to the Lagrangian density $%
\mathcal{L}_{\text{\textsc{f}}}$ is the usual one and we report it for
completeness:%
\begin{eqnarray*}
\frac{\partial \mathcal{L}_{\text{\textsc{f}}}}{\partial \left( \partial _{t}%
\mathbf{A}\right) }-\mathcal{L}_{\text{\textsc{f}}} &=&\left( \partial _{t}%
\mathbf{A+}\nabla \varphi \right) \cdot \partial _{t}\mathbf{A}-\frac{1}{2}%
\left( \partial _{t}\mathbf{A+}\nabla \varphi \right) ^{2}+\frac{1}{2}\left(
\nabla \times \mathbf{A}\right) ^{2} \\
&=&-\mathbf{E}\cdot \left( -\mathbf{E+}\nabla \varphi \right) -\frac{1}{2}%
\mathbf{E}^{2}+\frac{1}{2}\mathbf{H}^{2} \\
&=&\frac{1}{2}\mathbf{E}^{2}+\frac{1}{2}\mathbf{H}^{2}-\mathbf{E}\cdot
\nabla \varphi =\mathcal{E}_{\text{\textsc{f}}}-\mathbf{E}\cdot \nabla
\varphi
\end{eqnarray*}

\bigskip

The energy density relative to the Lagrangian density $\mathcal{L}_{\text{%
\textsc{i}}}$ is given by ()%
\begin{eqnarray*}
\frac{\partial \mathcal{L}_{\text{\textsc{i}}}}{\partial \left( \partial
_{t}S\right) }\cdot \partial _{t}S-\mathcal{L}_{\text{\textsc{i}}} &=&\beta
\varphi u\partial _{t}S-\beta \left( \mathbf{\QTR{mathbf}{A}}\cdot \nabla
S+\partial _{t}S\varphi \right) u \\
&=&-\beta \mathbf{\QTR{mathbf}{A}}\cdot \nabla Su=-\mathbf{\QTR{mathbf}{A}}%
\cdot \mathbf{\mathbf{j}}
\end{eqnarray*}%
If we set 
\begin{equation*}
\mathcal{E}_{\text{\textsc{i}}}=-\mathbf{E}\cdot \nabla \varphi -\mathbf{%
\mathbf{\QTR{mathbf}{A}}\cdot \mathbf{j}}
\end{equation*}%
we have that%
\begin{eqnarray*}
E_{\text{\textsc{i}}} &=&\int \mathcal{E}_{\text{\textsc{i}}}dx=\int \left( -%
\mathbf{E}\cdot \nabla \varphi -\mathbf{\mathbf{\QTR{mathbf}{A}}\cdot 
\mathbf{j}}\right) dx \\
&=&\int \left( \nabla \cdot \mathbf{E\ }\varphi -\mathbf{\mathbf{%
\QTR{mathbf}{A}}\cdot \mathbf{j}}\right) dx=\int \left( \rho \varphi -%
\mathbf{\mathbf{\QTR{mathbf}{A}}\cdot \mathbf{j}}\right) dx \\
&=&-\beta \int \left( \varphi \partial _{t}S+\mathbf{\mathbf{\QTR{mathbf}{A}}%
}\cdot \nabla S\right) udx
\end{eqnarray*}

Then 
\begin{equation*}
E=E_{\text{\textsc{m}}}+E_{\text{\textsc{f}}}+E_{\text{\textsc{i}}}.
\end{equation*}

$\square $

\bigskip

\textbf{Conservation of Momentum. }%
\index{momentum}\textbf{\ }Momentum, by definition, is the quantity which is
preserved by virtue of the space invariance of the Lagrangian. Here we will
compute only the matter-field moment since it is the only part needed it the
rest of this paper.

\begin{proposition}
\label{mome}The momentum takes the following form%
\begin{equation}
\mathbf{P}=\mathbf{P}_{%
\text{\textsc{m}}}+\mathbf{P}_{\text{\textsc{f}}}+\mathbf{P}_{\text{\textsc{i%
}}}
\end{equation}%
where 
\begin{eqnarray}
\mathbf{P}_{\text{\textsc{m}}} &=&\int \left( \partial _{t}u\nabla
u+\partial _{t}S\nabla Su^{2}\right) dx \\
&=&\int \left( \partial _{t}u\nabla u+\rho \mathbf{j}\right) dx
\end{eqnarray}%
is the \textbf{matter field} momentum.
\end{proposition}

\textbf{Proof}: By Noether's Theorem, we have that the momentum densities $%
\mathcal{P}_{\text{\textsc{m}}}$ relative to the Lagrangian densities $%
\mathcal{L}_{\text{\textsc{m}}}$ is given by 
\begin{equation*}
\mathcal{P}_{\text{\textsc{m}}}=\frac{\partial \mathcal{L}_{\text{\textsc{m}}%
}}{\partial \left( \partial _{t}u\right) }\nabla u+\frac{\partial \mathcal{L}%
_{\text{\textsc{m}}}}{\partial \left( \partial _{t}S\right) }\nabla
S=\partial _{t}u\nabla u+\partial _{t}S\nabla Su^{2}
\end{equation*}%
$\square $

\bigskip

\textbf{Conservation of electric charge}: Even if our equations are not
invariant for the whole gauge group, nevertheless the electric charge is
preserved as it is a consequence of (\ref{gau}). In fact by (\ref{gauss})
and (\ref{amp}), we get the continuity equation: 
\begin{equation*}
\partial _{t}\rho =\nabla \cdot \left( \partial _{t}\mathbf{E}\right)
=\nabla \cdot \left( \nabla \times \mathbf{H-j}\right) =-\nabla \cdot 
\mathbf{j}
\end{equation*}%
So, the total electric charge 
\begin{equation}
Q\left[ u,S\right] =\int \rho dx=-\beta \int \partial _{t}Su\ dx  \label{ccc}
\end{equation}%
is preserved.

\bigskip

\textbf{Conservation of hylenic charge}: Following \cite{milano} and \cite%
{befolibro} the hylenic charge, by definition, is the quantity which is
preserved by the invariance for the tranformation (\ref{trg}). It is defined
as follows:%
\begin{equation}
H\left[ u,S,\varphi \right] =\int \left( \frac{\partial _{t}Su^{2}}{\beta }%
-\varphi u\right) dx=\int \left( \rho -\varphi \right) u\ dx  \label{hc}
\end{equation}

By equation (\ref{e2}), we see directly that the hylenic charge is preserved.

\subsection{The Cauchy problem}

In order to study the Cauchy problem, it is more convenient to use the
variable $\psi $ rather that $(u,S).$ To this end, we introduce the
following operators:%
\begin{eqnarray*}
\mathfrak{D}_{t}\left( \psi \right) &=&\func{Im}\left( \frac{\partial
_{t}\psi }{\psi }\right) =\func{Im}\left( \frac{\partial _{t}\left(
ue^{iS}\right) }{ue^{iS}}\right) \\
&=&\func{Im}\left( \frac{\partial _{t}ue^{iS}+iu\partial _{t}Se^{iS}}{ue^{iS}%
}\right) =\func{Im}\left( \partial _{t}u+i\partial _{t}S\right) =\partial
_{t}S
\end{eqnarray*}%
and 
\begin{equation*}
\mathfrak{D}_{x}\left( \psi \right) =\func{Im}\left( \frac{\nabla \psi }{%
\psi }\right) =\nabla S
\end{equation*}%
Since we have assumed the Lorentz condition the equations (\ref{e1}),...,(%
\ref{e4}), using (\ref{WN}), can be rewritten as follows:%
\begin{equation}
\square \psi +\psi =-N^{\prime }(\left\vert \psi \right\vert )\frac{\psi }{%
\left\vert \psi \right\vert }+\mathbf{\QTR{mathbf}{A}}\cdot \mathfrak{D}%
_{x}\left( \psi \right) -\varphi \mathfrak{D}_{t}\left( \psi \right)
\label{e1+}
\end{equation}%
\begin{equation}
\square \varphi =\mathfrak{D}_{t}\left( \psi \right) \left\vert \psi
\right\vert  \label{e3+}
\end{equation}%
\begin{equation}
\square \mathbf{A}=\mathfrak{D}_{x}\left( \psi \right) \left\vert \psi
\right\vert .  \label{e4+}
\end{equation}

We make the following (redundant) assumptions on $N:$%
\begin{equation}
N,\ N^{\prime }\ \text{and\ }N^{\prime \prime }\ \text{are\ bounded;}
\label{aa}
\end{equation}%
\begin{equation}
N\left( s\right) \geq -\frac{1}{2}\left( 1-\delta \right) s^{2};\ 0<\delta
<1.  \label{bbb}
\end{equation}

\begin{theorem}
If (\ref{aa}), (\ref{bbb}) hold and $\beta $ is sufficiently small, the
Cauchy problem relative to equations (\ref{e1+}),...,(\ref{e4+}) has a
unique weak solution.
\end{theorem}

\textbf{Proof}: The proof of this theorem follows standard arguments and we
will just give a sketch. The function space where to work is $H^{1}\times
\left( \mathcal{D}^{1,2}\right) ^{4}$ where%
\begin{equation*}
H^{1}=\left\{ \psi \in L^{2}(\mathbb{R}^{3},\mathbb{C})\ |\ \int \left(
\left\vert \nabla \psi \right\vert ^{2}+\left\vert \psi \right\vert
^{2}\right) dx<+\infty \right\} 
\end{equation*}%
\begin{equation}
\mathcal{D}^{1,2}=\left\{ f\in L^{6}(\mathbb{R}^{3})\ |\ \int \left\vert
\nabla f\right\vert ^{2}dx<+\infty \right\} .  \label{D}
\end{equation}%
We set

\begin{equation*}
U\left( t,x\right) =(\psi (t,x),\varphi (t,x),\mathbf{A}(t,x));
\end{equation*}%
where $\psi \in H^{1}(\mathbb{R}^{3},\mathbb{C}),$ $\varphi \in \mathcal{D}%
^{1,2}(\mathbb{R}^{3},\mathbb{R}),$ and $\mathbf{A}\in \mathcal{D}^{1,2}(%
\mathbb{R}^{3},\mathbb{R}^{3})=\left[ \mathcal{D}^{1,2}(\mathbb{R}^{3},%
\mathbb{R}^{1})\right] ^{3}.$ So, we end with the Cauchy problem%
\begin{equation}
\square U+P_{1}U=F(U)  \label{U}
\end{equation}%
where $P_{1}$ is the projection of $U$ on the first component i.e. $%
P_{1}U=\psi $.

We equip the phase space $X:=\left[ H^{1}\times \left( \mathcal{D}%
^{1,2}\right) ^{4}\right] \times \left( L^{2}\right) ^{6},$ with its natural
norm given by%
\begin{equation*}
\left\Vert U\right\Vert ^{2}=\left\Vert \partial _{t}\psi \right\Vert
_{L^{2}}^{2}+\left\Vert \partial _{t}\psi \right\Vert
_{H^{1}}^{2}+\left\Vert \partial _{t}\varphi \right\Vert
_{L^{2}}^{2}+\left\Vert \nabla \varphi \right\Vert _{L^{2}}^{2}+\left\Vert
\partial _{t}\mathbf{A}\right\Vert _{L^{2}}^{2}+\left\Vert \nabla \mathbf{A}%
\right\Vert _{L^{2}}^{2}
\end{equation*}%
It is well known that a sufficient condition for the Cauchy problem to have
a unique solution for the initial data in $X$ is:

\begin{itemize}
\item the energy inequality holds: there exists two positive constants $%
c_{1} $ and $c_{2}$ such that%
\begin{equation*}
c_{1}\left\Vert U\right\Vert ^{2}\leq E\left[ U\right] \leq c_{2}\left\Vert
U\right\Vert ^{2}
\end{equation*}%
This inequality can be proved if $\beta $ is sufficiently small and if (\ref%
{aa}) holds;

\item $F:X\rightarrow X^{\prime }$ is locally compact; this fact holds since
the embedding 
\begin{equation*}
U_{loc}\rightarrow \left( L_{loc}^{6}\right) ^{6}
\end{equation*}%
is compact;
\end{itemize}

Under these conditions the proof goes as follows:

\begin{enumerate}
\item we take a sequence of approximate solutions; for example we can use
the Faedo-Galerkin procedure;

\item we take the weak limit of the approximated solutions which exists
thank to the second energy inequality;

\item we pass to the limit in the weak formulation of the equations; we can
take the limit in the nonlinear part $F$ since it is locally compact;

\item we can prove the uniqueness thanks to the first energy inequality and
the Gronwall's inequality.
\end{enumerate}

$\square $

\begin{remark}
The optimal conditions for the existence of solutions and the study of the
their regularity is not in the aim of this paper and it is a question that,
for the moment, is left open.
\end{remark}

\section{ $q$-solitons\label{qs}}

Roughly speaking a \textit{solitary wave} is a solution of a field equation
whose energy travels as a localized packet and which preserves this
localization in time. A \textit{soliton} is a solitary wave which exhibits
some form of stability so that it has a particle-like behavior (see e.g. 
\cite{dodd},\cite{yang},\cite{raj},\cite{befolibro}).\ 

It is well known that equation (\ref{NKG}) presents solitons under suitable
assumptions on $W$. It\ has been largely studied during the 70's and the
80's of the last century. The first rigorous result about finite energy
solution was due to Strauss \cite{strauss} and later Beresticky and Lions 
\cite{Beres-Lions} gave sufficient and "almost necessary" condition for the
existence. In \cite{befolibro} there is a detailed analysis of the case in
which $W\geq 0$. If we couple (\ref{NKG}) with the Maxwell equation via the
interaction (\ref{W}) the solitons usually are called $Q$-balls (Coleman 
\cite{Coleman86}). The first rigorous result about the existence of $Q$%
-balls has been establish in 2002, \cite{befo2002}. Afterwards, their
stability has been proved in \cite{befo11max}. A detailed analysis on $Q$%
-balls and the references to the large literature can be found in \cite%
{befolibro}; in all these paper the interaction between the solitons and the
e.m. field is established by the Lagrangian (\ref{W}).

In this section we analyze the existence and the properties of solitons when
the interaction with the Maxwell equation is simply given by the Lagrangian (%
\ref{Li}) and not by (\ref{W}). They will be called $q$-solitons. The main
difference between $q$-solitons and the $Q$-balls is that the former behave
like single particles while the latter behave like a swarm of particles (see 
\cite{befolibro}, Sections 4.1.2 and 5.1.5).

\subsection{Existence of stationary waves}

Let us prove the existence of some particular solution of equations (\ref{e1}%
),...,(\ref{e4}); first, we look for stationary solutions, namely solutions
where $\psi $ is a stationary wave, i.e. 
\begin{equation}
\psi (t,x)=u(x)e^{-i\omega t};\ u\geq 0  \label{SW}
\end{equation}%
We make the following ansaz:%
\begin{equation}
u=u(x);\ \ S=-\omega t;\ \ \varphi =\varphi (x);\ \ \mathbf{\QTR{mathbf}{A}}%
=0.  \label{SW+}
\end{equation}%
Replacing these variables in (\ref{e1}),...,(\ref{e4}), we have that eq. (%
\ref{e2}) and (\ref{e4}) are identically satisfied while eq. (\ref{e1}) and (%
\ref{e3}) become

\begin{equation}
-\Delta u+W^{\prime }(u)-\omega ^{2}u+\beta \omega \varphi =0  \label{KG}
\end{equation}%
\begin{equation}
-\Delta \varphi =\beta \omega u  \label{poisson}
\end{equation}

These two equations have nontrivial solutions provided that suitable
conditions on $W\in C^{2}$ be satisfied: we write $W$ as follows,%
\begin{equation}
W(s)=\frac{1}{2}s^{2}+N(s),\   \label{NN}
\end{equation}%
In the model of our interest, $N$ most be considered as a small perturbation
of the parabola $1/2s^{2}.$ However, in order to get an existence result ,
it is sufficient to make the following assumptions on $N:$

\begin{itemize}
\item (N-1)$\ N(0)=N^{\prime }(0)=N^{\prime \prime }(0)=0;$

\item (N-2) $\inf_{s\in \mathbb{R}^{+}}N(s):=N_{\inf }<0,$ ($N_{\inf }$ is
allowed also to be $-\infty $);

\item (N-3) there exist $C>0$ and $2<p<6$ such that $\left\vert N^{\prime
}(s)\right\vert \leq C(1+s^{p-1}).$
\end{itemize}

We will show that, at least for $\beta \ $small, the above assumptions
guarantee the existence of nontrivial solutions of eqs. (\ref{KG}), (\ref%
{poisson}). In most of the literature relative to (\ref{NKG}) usually we
have the following choice of $N:$%
\begin{equation}
N(s)=\frac{1}{p}\left\vert s\right\vert ^{p},\ 2<p<6.  \label{power}
\end{equation}%
This assumption implies the existence of nontrivial solutions also for eqs. (%
\ref{KG}), (\ref{poisson}) for every $\beta >0$. However, in our model, it
is more interesting (see Th. \ref{lilli}) to choose a "bump-like" $N$ such
as 
\begin{equation}
N(s)=-\varepsilon ^{2}s^{3}\exp \left( -\left\vert \frac{s-1}{\varepsilon }%
\right\vert ^{2}\right) .  \label{bump}
\end{equation}%
or a "bell" function such as%
\begin{equation}
N(s)=\left\{ 
\begin{array}{cc}
-\left[ \left( s-1\right) ^{2}-\varepsilon ^{2}\right] ^{2} & 
if\;\;|s-1|<\varepsilon \\ 
0 & if\;\;\left\vert s-1\right\vert \geq \varepsilon%
\end{array}%
\right.  \label{bell}
\end{equation}%
where $\varepsilon $ is a small parameter which makes $W(s)\geq 0.$ Its
relevance will be discussed in Th. \ref{lilli}.

We define the following bilinear form:%
\begin{equation}
a_{\omega }(u,u)=\frac{1}{2}\int \left[ \left\vert \nabla u\right\vert
^{2}+\left( 1-\omega ^{2}\right) u^{2}\right] dx+\beta ^{2}\omega ^{2}\iint 
\frac{u(x)u(y)}{\left\vert x-y\right\vert ^{2}}dxdy  \label{a}
\end{equation}%
Notice that 
\begin{equation*}
\iint \frac{u(x)u(y)}{\left\vert x-y\right\vert ^{2}}dxdy=\int \left( G\ast
u\right) u\ dx
\end{equation*}%
where%
\begin{equation*}
G(x)=\frac{1}{4\pi \left\vert x\right\vert ^{2}}
\end{equation*}%
is the Green function relative to the Poisson equation%
\begin{equation}
-\Delta \phi =u  \label{poisson+}
\end{equation}%
namely $\left( G\ast u\right) (x)=\left( -\Delta \right) ^{-1}:\left( 
\mathcal{D}^{1,2}\right) \rightarrow \mathcal{D}^{1,2}$.

Now, let us introduce a number $\omega _{\inf }$ which is very relevant in
this study of solitons:%
\begin{equation}
\omega _{\inf }:=\inf \ \left\{ \omega >0\ |\ \exists u\in H_{c}^{1},\
a_{\omega }(u,u)+\int N(u)\ dx<0\right\} ;  \label{oinf}
\end{equation}%
$\omega _{\inf }$ depends on $\beta $ and the shape of $N.$ For example if $%
N(u)$ is given by (\ref{power}), then it is immediate to check that $\omega
_{\inf }=0$ for every $\beta >0$. If $N(u)$ is given by (\ref{bump}) $\omega
_{\inf }$ depends on $\beta $ (see Cor. \ref{lilli+}).

We have the following theorem.

\begin{theorem}
\label{lillo}If (N-1) and (N-3) hold and if $\omega _{\inf }<1$, then for
every $\omega \in \left( \omega _{\inf },1\right) ,$ eqs. (\ref{KG}), (\ref%
{poisson}) have nontrivial solutions in $H^{1}(\mathbb{R}^{3})\times 
\mathcal{D}^{1,2}(\mathbb{R}^{3})$.
\end{theorem}

\textbf{Proof}: The couple of equations (\ref{KG}) and (\ref{poisson}) can
be easily solved by standard variational methods; we will give here a sketch
of the proof avoiding standard estimates which are well known among people
working in nonlinear analysis.

Set 
\begin{equation*}
H_{\text{rad}}^{1}\left( \mathbb{R}^{3}\right) =\left\{ u\in H^{1}\left( 
\mathbb{R}^{3}\right) \ |\ u=u(\left\vert x\right\vert )\right\} ;
\end{equation*}%
and 
\begin{equation*}
V:=\left\{ u\in H_{\text{rad}}^{1}\left( \mathbb{R}^{3}\right) \ |\
a_{\omega }(u,u)<+\infty \right\}
\end{equation*}%
Since $\omega _{\sup }<1,$ $V$ is a Hilbert space equipped with scalar
product $a_{\omega }(u,v)$ and norm $\left\Vert u\right\Vert _{V}:=\sqrt{%
a_{\omega }(u,v)}.$ Moreover, by the definition of $\omega _{\sup },$ $%
\exists \delta >0,$%
\begin{equation*}
\left\Vert u\right\Vert _{V}^{2}\geq \delta \left\Vert u\right\Vert
_{H^{1}}^{2}
\end{equation*}%
Then, using the Gagliardo-Nirenberg-Sobolev estimate we can see that%
\begin{equation}
V=H_{\text{rad}}^{1}+\left[ \mathcal{D}_{\text{rad}}^{1,2}(\mathbb{R}^{3})%
\right] ^{\prime }\subset H_{\text{rad}}^{1}+L^{6/5}  \label{V}
\end{equation}%
Now, we define on $V$ the following functional:%
\begin{eqnarray}
J\left[ u\right] &=&\left\Vert u\right\Vert _{V}^{2}+\int N(u)dx  \label{J}
\\
&=&\frac{1}{2}\int \left[ \left\vert \nabla u\right\vert ^{2}+\left(
1-\omega ^{2}\right) u^{2}\right] \ dx+\omega ^{2}\beta ^{2}\int \left[
\left( G\ast u\right) u+N(u)\right] dx  \notag
\end{eqnarray}%
By (N-3), (\ref{V}) and standard arguments, $J$ is a differentiable
functional in $V.$ So we have to prove two facts: (1) the critical points of 
$J$ solve eqs. (\ref{KG}) and (\ref{poisson}) and (2) if $\omega \in \left(
\omega _{\inf },\ 1\right) ,$ $J$ has at least a nontrivial critical point.

(1) We have that $\forall v\in V$ 
\begin{equation*}
dJ\left[ u\right] (v)=\int \left[ \nabla u\nabla v+\left( 1-\omega
^{2}\right) uv\right] \ dx+\omega ^{2}\beta ^{2}\int \left[ \left( G\ast
u\right) +N^{\prime }(u)\right] vdx
\end{equation*}%
and by well known arguments, we have that

\begin{equation*}
-\Delta u+(1-\omega ^{2})u+\omega ^{2}\beta ^{2}\left( G\ast u\right)
+N^{\prime }(u)=0.
\end{equation*}%
Taking account of (\ref{NN})%
\begin{equation*}
-\Delta u+W^{\prime }(u)+\omega ^{2}\beta ^{2}\left( G\ast u\right) -\omega
^{2}u=0;
\end{equation*}%
finally, setting $\varphi =\omega \beta \left( G\ast u\right) ,$ we get
equation (\ref{KG}) while equation (\ref{poisson}) follows from the
definition of $\varphi .$

(2) The simplest way to prove the existence of critical points of $J$ is the
use of the Mountain Pass theorem of Ambrosetti and Rabinowitz \cite{AR}.
Following standard arguments, it is easy to prove that $J$ satisfies the
Palais-Smale condition (for a very similar result see \cite{befo2002}, Lemma
4.3). The interesting fact is to check the conditions which guarantee the
geometry of the Mountain Pass theorem namely that%
\begin{equation}
\exists r>0,\ \left\Vert u\right\Vert _{V}^{2}=r\Rightarrow J\left[ u\right]
\geq b>0  \label{g1}
\end{equation}%
and%
\begin{equation}
\exists \bar{u},\ \left\Vert \bar{u}\right\Vert _{H^{1}}^{2}>r,\ J\left[ 
\bar{u}\right] \leq 0  \label{g2}
\end{equation}%
By the definition of $\left\Vert \cdot \right\Vert _{V}$ 
\begin{equation*}
\left\Vert u\right\Vert _{V}^{2}\geq \delta \left\Vert u\right\Vert
_{H^{1}}^{2},\ \ \delta >0.\ 
\end{equation*}%
If $r>0,$ is sufficiently small, by (\ref{NN}), (N-1), (N-3) and standard
computations, $\exists C,\eta >0$ such that 
\begin{equation*}
\int \left\vert N(u)\right\vert dx\leq C\left\Vert u\right\Vert
_{H^{1}}^{2+\eta }\leq \frac{C}{\delta }\left\Vert u\right\Vert
_{H^{1}}^{2+\eta }
\end{equation*}%
then, if $\left\Vert u\right\Vert _{V}=r$ 
\begin{eqnarray*}
J\left[ u\right] &=&\frac{1}{2}\left\Vert u\right\Vert _{V}^{2}+\int
\left\vert N(u)\right\vert dx \\
&\geq &\delta \left\Vert u\right\Vert _{V}^{2}-\frac{C}{\delta }\left\Vert
u\right\Vert _{V}^{2+\eta } \\
&=&\left[ \delta -\frac{Cr^{\eta }}{\delta }\right] r^{2}
\end{eqnarray*}%
Then if $r$ is sufficiently small $J\left[ u\right] \geq b>0$ and (\ref{g1})
is proved. (\ref{g2}) holds by the definition (\ref{oinf}) of $\omega _{\inf
}$.

$\square $

\begin{corollary}
\label{lilli+}If (N-1), (N-2) and (N-3) hold, then there exixts $\beta
_{0}>0 $, such that for every $\beta \in \left( 0,\beta _{0}\right) $ eqs. (%
\ref{KG}), (\ref{poisson}) have nontrivial solutions in $H^{1}(\mathbb{R}%
^{N})$.
\end{corollary}

\textbf{Proof: }By Th. \ref{lillo}, it is sufficient to prove that%
\begin{equation*}
\omega _{\inf }<1.
\end{equation*}%
By (N-2), we can choose a point $s_{1}$ such that%
\begin{equation*}
N(s_{1})=-h^{2}.
\end{equation*}%
We set%
\begin{equation}
u_{r}=\left\{ 
\begin{array}{cc}
s_{1} & if\;\;|x|<r \\ 
0 & if\;\;|x|>r+1 \\ 
\frac{|x|}{r}s_{1}-\left[ \left( r+1\right) \left\vert x\right\vert -1\right]
s_{1} & if\;\;r<|x|<r+1%
\end{array}%
\right. ,  \label{inff}
\end{equation}%
\begin{equation*}
\bar{\omega}=\sqrt{1-\frac{h^{2}}{s_{1}^{2}}},
\end{equation*}%
and%
\begin{equation*}
F\left[ u\right] :=\frac{1}{2}\int \left\vert \nabla u\right\vert
^{2}dx+\int \left( W(u)-\frac{1}{2}\bar{\omega}^{2}u^{2}\right) dx
\end{equation*}

Let us compute $F\left[ u_{r}\right] :$ 
\begin{eqnarray*}
F\left[ u_{r}\right] &=&\frac{1}{2}\int_{B_{r+1}\backslash B_{r}}\left\vert
\nabla u_{r}\right\vert ^{2}dx+\int_{B_{r+1}\backslash B_{r}}\left( W(u)-%
\frac{1}{2}\bar{\omega}^{2}u_{r}^{2}\right) dx \\
&&+\int_{B_{r}}\left( W(u_{r})-\frac{1}{2}\bar{\omega}^{2}u_{r}^{2}\right) dx
\end{eqnarray*}%
The first part can be estimated as follows: 
\begin{eqnarray*}
\frac{1}{2}\int_{B_{r+1}\backslash B_{r}}\left\vert \nabla u_{r}\right\vert
^{2}dx+\int_{B_{r+1}\backslash B_{r}}\left( W(u)-\frac{1}{2}\bar{\omega}%
^{2}u^{2}\right) dx &\leq &C\cdot meas\left( B_{r+1}\backslash B_{r}\right)
\\
&\leq &C_{1}r^{2}
\end{eqnarray*}%
For the second part we have that%
\begin{eqnarray*}
\int_{B_{r}}\left( W(u_{r})-\frac{1}{2}\bar{\omega}^{2}u_{r}^{2}\right) dx
&=&\int_{B_{r}}\left[ \frac{1}{2}u_{r}^{2}-\frac{1}{2}\bar{\omega}%
^{2}u_{r}^{2}+N(u_{r})\right] dx \\
&=&\int_{B_{r}}\left[ \frac{1}{2}\left( 1-\bar{\omega}^{2}\right)
s_{1}^{2}+N(u_{r}\left( s_{1}\right) )\right] dx \\
&\leq &\int_{B_{r}}\left[ \frac{1}{2}\left[ 1-\left( 1-\frac{h^{2}}{s_{1}^{2}%
}\right) \right] s_{1}^{2}-h^{2}\right] dx \\
&=&\int_{B_{r}}\left[ \frac{1}{2}h^{2}-h^{2}\right] dx=\frac{4}{3}\pi r^{3}%
\frac{h^{2}}{2}=\frac{2}{3}\pi r^{3}h^{2}
\end{eqnarray*}%
Then, we have that%
\begin{equation*}
F\left[ u_{r}\right] \leq C_{1}r^{2}-\frac{2}{3}\pi r^{3}h^{2}.
\end{equation*}%
So, we can choose $\bar{r}$ so large that%
\begin{equation*}
F\left[ u_{\bar{r}}\right] <-\bar{r}^{3}h^{2}<-1
\end{equation*}%
and $\beta $ so small that%
\begin{equation*}
\beta ^{2}\bar{\omega}^{2}\int \left( G\ast u_{\bar{r}}\right) u_{\bar{r}%
}dx\leq 1.
\end{equation*}%
Then 
\begin{equation*}
a_{\bar{\omega}}(u_{\bar{r}},u_{\bar{r}})+\int N(u_{\bar{r}})\ dx=F\left[ u_{%
\bar{r}}\right] +\frac{1}{2}\beta ^{2}\bar{\omega}^{2}\int \left( G\ast u_{%
\bar{r}}\right) u_{\bar{r}}dx\leq -\ \frac{1}{2}
\end{equation*}%
and hence 
\begin{equation*}
\omega _{\inf }<\bar{\omega}=\sqrt{1-\frac{h^{2}}{s_{1}^{2}}}<1.
\end{equation*}

$\square $

\subsection{Stationary $q$-solitons\label{ss}}

Using the equivariant Mountain Pass theorem and exploiting the fact that the
functional (\ref{J}) is even, it is possible to prove that eqs. (\ref{e1}%
),...,(\ref{e4}), have infinitely many radially symmetric solutions of the
form (\ref{SW}), (\ref{SW+}), namely solitary waves. We call \textbf{ground
state solution}, the radially symmetric solution $u_{0}>0$ which minimizes
the following quantity:%
\begin{equation*}
\Lambda \left[ u,\omega \right] =\frac{E\left[ u,\omega \right] }{\left\vert
H\left[ u,\omega \right] \right\vert }=\frac{\int \left( \frac{1}{2}%
\left\vert \nabla u\right\vert -\frac{1}{2}\omega ^{2}u^{2}+W(u)+\beta
^{2}\omega ^{2}\left( G\ast u\right) \right) dx}{\left\vert \int \left(
-\beta ^{-2}\omega u^{2}-\omega \left( G\ast u\right) u\right) dx\right\vert 
}
\end{equation*}%
in $H^{1}\times \mathbb{R}^{+}.$ Clearly, at least for a generic $W$, this
solution is unique and it corresponds to the critical value determined by
the Mountain Pass Theorem, having chosen $\omega $ which minimizes $\Lambda %
\left[ u,\omega \right] $. Notice that $\Lambda \left[ u,\omega \right] $ is
the ratio of the matter energy (\ref{me}) and the hylenic charge (\ref{hc}).
If $W\geq 0,$ the ground state solution, is a soliton in the sense that it
is orbitally stable (see e.g. \cite{befo11max} or \cite{befolibro}).

\bigskip

From now on, $\sigma _{0}$ we will denote the ground state solution of
equations (\ref{KG}),(\ref{poisson}), namely the configuration

\begin{equation}
\sigma _{0}(x)=\left[ 
\begin{array}{c}
(u_{0}(x),0,0,0) \\ 
(0,-\omega _{0}+\theta ,0,0)%
\end{array}%
\right]  \label{q-s}
\end{equation}%
where $\theta $ is a possible phase shift which is not relevant and from now
on, it will be neglected. Such a function will be called a $q$\textbf{%
-soliton}. We have chosen this name to emphasize the comparison with the $Q$%
-balls that are stable configurations of the equations determined by the
action $\mathbb{A}_{\text{\textsc{w}}}+\mathbb{A}_{\text{\textsc{f}}}$ (see\
(\ref{W}) and (\ref{1}) and the discussion at the beginning of Sec. \ref{qs}%
). Roughly speaking a $Q$-ball behaves like a swarm of charged particles
kept close to each other by the gluing force determined by $N(s)$ (see \cite%
{milano} or \cite{befolibro} Sec. 5.1.5). Instead, as we will see in this
and the next sections, a $q$\textbf{-soliton }behaves like a single particle
of "matter" condensed by the gluing force determined by $N(s)$.

The $q$-soliton, has a positive electric charge $\rho _{0}=\omega u_{0}\ $%
(and hence, by (\ref{poisson}), $\varphi _{0}(x)>0$). However, the equations
(\ref{e1}),...,(\ref{e4}), have also a solution with negative charge given
by 
\begin{equation*}
\sigma _{0}^{-}(x)=\left[ 
\begin{array}{c}
(u_{0}(x),0,0,0) \\ 
(0,\omega _{0},0,0)%
\end{array}%
\right]
\end{equation*}%
Then,(\ref{e1}),..,(\ref{e4}), have at least two orbitally stable solutions
determined by a $q$-soliton $\sigma _{0}(x)$ and a $q$-antisoliton $\sigma
_{0}^{-}(x)$:%
\begin{equation*}
U(t,x)=\left[ 
\begin{array}{c}
(u_{0}(x),0,\varphi _{0}(x),0) \\ 
(0,-\omega _{0}t,0,0)%
\end{array}%
\right] ;\ U^{-}(t,x)=\left[ 
\begin{array}{c}
(u_{0}(x),0,-\varphi _{0}(x),0) \\ 
(0,\omega _{0}t,0,0)%
\end{array}%
\right]
\end{equation*}%
Generally, they are unique up to space-time translations and phase shift.
Rotations do not produce new solutions since $u_{0}$ is radially symmetric.

\bigskip

The "shape" of a soliton is determined by the nonlinear $N(s)$. In \cite%
{hylo} there is a detailed analysis of this topic in the case $\beta =0$.
Clearly this analysis can be extended to the $q$-soliton when $\beta $ is
small. The next theorem examines some properties of the $q$-solitons in the
case in which $N(s)$ is a small bump such as (\ref{bell}):

\begin{theorem}
\label{lilli}For every $\varepsilon >0,$ we can choose $N$ such that

\begin{itemize}
\item $W\geq 0;$

\item $1-\varepsilon <\omega _{\inf }<1.$

\item if $u_{0}$ is the Mountain Pass solution of eqs. (\ref{KG}) and (\ref%
{poisson}), then 
\begin{equation*}
1-\varepsilon \leq \left\Vert u_{0}\right\Vert _{L^{\infty }}\leq
1+\varepsilon
\end{equation*}
\end{itemize}
\end{theorem}

\textbf{Proof. }We choose $N$ to be a bell function such as (\ref{bell}) so
that%
\begin{equation*}
\min N=N\left( 1\right) \geq -\varepsilon ^{2}
\end{equation*}%
and 
\begin{equation*}
supp(N)=\left[ 1-\varepsilon ,\ 1+\varepsilon \right]
\end{equation*}%
Then, the first inequality is trivially verified. In order to prove the
second inequality, we see that%
\begin{equation*}
N(u)\geq -\frac{1}{2}\varepsilon ^{2}
\end{equation*}%
and so, if we put $\bar{\omega}^{2}=1-\varepsilon ^{2},$ we have that%
\begin{eqnarray*}
a_{\bar{\omega}}(u,u)+\int N(u)\ dx &\geq &\frac{1}{2}\int \left[ \left\vert
\nabla u\right\vert ^{2}+\left( 1-\bar{\omega}^{2}\right) u^{2}\right] \
dx+\int N(u)dx \\
&=&\frac{1}{2}\int \left[ \left\vert \nabla u\right\vert ^{2}+\varepsilon
^{2}u^{2}\right] \ dx-\frac{1}{2}\varepsilon ^{2}\int u^{2}dx\geq 0
\end{eqnarray*}
So by the definition of $\omega _{\inf }$ (\ref{oinf}), we have that $\omega
_{\inf }^{2}>\bar{\omega}^{2}$ and hence $\omega _{\inf }>\sqrt{%
1-\varepsilon ^{2}}>1-\varepsilon $.

The third inequality follows applying to eq. (\ref{KG}) the maximum
principle. The details of the proof can bo found in \cite{hylo}.

$\square $

\bigskip

\begin{remark}
\label{R}The picture which comes out from Cor. \ref{lilli+} and Th. \ref%
{lilli} is the following: given the free electromagnetic field and the free
matter field relative to KG, we get $q$-solitons provided that

\begin{itemize}
\item the interaction between them is given by a Lagrangian of type (\ref{Li}%
) with $\beta $ \textbf{very small};

\item KG is perturbed by a nonlinear term $N(s)$ (negative in some point) 
\textbf{small} with respect to $1$ and \textbf{large} with respect to $\beta
.$
\end{itemize}
\end{remark}

\bigskip

If we want to analyze the properties of a $q$-soliton considered as a model
for physical particles, it is useful to rewrite equation (\ref{e1}) with
dimensional constants. We get the following equation, (which is satisfied by 
$\sigma _{0}$ if suitably rescaled):%
\begin{equation}
\partial _{t}^{2}u-c^{2}\Delta u+\alpha ^{2}u+\frac{c^{2}}{\ell ^{2}}%
N^{\prime }(u)-\left[ \left( \partial _{t}S\right) ^{2}-c^{2}\left\vert
\nabla S\right\vert ^{2}\right] u=\eta \beta \left( \mathbf{\QTR{mathbf}{A}}%
\cdot \nabla S-\varphi \partial _{t}S\right)  \label{dim}
\end{equation}

In this equation,

\begin{itemize}
\item $c$ is the speed of light which makes the equation invariant for the
Lorentz transformations with the parameter $c$;

\item $u$ has the dimension of 
\begin{equation*}
\frac{\left\{ \text{mass}\right\} ^{\frac{1}{2}}}{\left\{ \text{space}%
\right\} };
\end{equation*}%
this fact can be deduced e.g. by the fact that, by Th. \ref{E}, 
\begin{equation*}
\frac{1}{2}\int \left[ \left\vert \partial _{t}u\right\vert
^{2}+c^{2}\left\vert \nabla u\right\vert ^{2}\right] dx
\end{equation*}%
has the dimension of energy;

\item $\alpha $ has the dimension of a frequency; if we linearize eq. (\ref%
{dim}) with $\beta =0$, we get KG 
\begin{equation}
\partial _{t}^{2}u-c^{2}\partial _{x}^{2}u+\alpha ^{2}u=0.
\end{equation}%
which has the following dispersion relations:%
\begin{equation*}
\omega _{\text{\textsc{kg}}}=\alpha \sqrt{1+\frac{c^{2}}{\alpha ^{2}}k_{%
\text{\textsc{kg}}}^{2}}
\end{equation*}%
where $\omega _{\text{\textsc{kg}}}$ and $k_{\text{\textsc{kg}}}$ are the
frequency and the wave number of the small perturbations of the matter
field. Since $\omega _{0}<\alpha <\omega _{\text{\textsc{kg}}}$ the
oscillations of the $q$-soliton, having frequency $\omega _{0}$, do not
excite dispersive waves in the surrounding matter field. This fact partially
explains the stability of the soliton;

\item if we give to $N^{\prime }\ $the same dimension of $u$, $\ell $ has
the dimension of a length and it is of the order of the radius of the
soliton in the sense that 
\begin{equation*}
u_{0}(x),\ \nabla u_{0}(x)\cong 0\ \ \text{for every }x\geq r_{0}:=k\ell
\end{equation*}%
where $\cong $ means that the quantity is exponentially small and $k$ is a
dimensionless variable which depends on $N.$

\item here $S$ is supposed to be dimensionless;

\item $\eta \beta $ represents the strength of the interaction of the matter
field with the electromagnetic field; by (\ref{ecar}) 
\begin{equation*}
\dim \beta =\frac{\left\{ \text{electric}\ \text{charge}\right\} \cdot
\left\{ \text{time}\right\} \cdot \left\{ \text{space}\right\} }{\left\{ 
\text{mass}\right\} ^{\frac{1}{2}}}
\end{equation*}%
and if we give to $\varphi $ the dimension of an electric field i.e. 
\begin{equation*}
\left\{ \text{electric}\ \text{charge}\right\} /\left\{ \text{space}\right\}
,
\end{equation*}
then 
\begin{equation*}
\dim \eta =\frac{\left\{ \text{mass}\right\} }{\left\{ \text{electric}\ 
\text{charge}\right\} ^{2}\cdot \left\{ \text{space}\right\} ^{2}\cdot
\left\{ \text{time}\right\} ^{2}};
\end{equation*}%
using these variables, then $a_{\omega }(u,u)$ defined by (\ref{a}) becomes 
\begin{equation}
a_{\omega }(u,u)=\int \left[ c^{2}\left\vert \nabla u\right\vert ^{2}+\left(
\alpha ^{2}-\omega ^{2}\right) u^{2}\right] dx+\omega ^{2}\eta ^{2}\beta
^{2}\iint \frac{u(x)u(y)}{\left\vert x-y\right\vert ^{2}}dxdy;  \label{a+}
\end{equation}%
hence, if $\eta \beta $ is too large with respect to the other constants,
then, by(\ref{oinf}), if $\omega _{\inf }\geq \alpha $ and there are no
solitons. Actually there is a competition between the gluing force which
increases with $c\ell ^{-1}N$ and the electric force which increases with $%
\eta \beta .$ The gluing force tends to concentrate the matter field while
the electric force tends to spread it.
\end{itemize}

\bigskip

By this discussion, it results that 
\begin{equation*}
W(u)=\frac{1}{2}\alpha ^{2}u+\frac{c^{2}}{\ell ^{2}}N(u)
\end{equation*}%
represents the potential of the "nuclear force" which is repelling when $u$
is small and attractive when the values of $u$ are in range where $N(s)$ is
negative. $N(s)$ is responsible of the nonlinear behavior of the matter
field and hence of the existence of $q$-solitons. By these considerations,
Th. \ref{lilli} and Remark \ref{R}, a $q$-soliton is a good model for
physical particles if, in the dimensionless equation 
\begin{equation*}
\beta ^{2}\ll \max N(s)\ll 1
\end{equation*}%
Also the condition 
\begin{equation}
W(u)\geq 0  \label{wlo}
\end{equation}%
is suitable for a physical model.

\bigskip

We denote by%
\begin{equation}
\left\{ \sigma _{0},\varphi _{0}\right\} =\left[ 
\begin{array}{c}
(u_{0}(x),0,\varphi _{0},0) \\ 
(0,-\omega _{0},0,0)%
\end{array}%
\right]
\end{equation}%
the equilibrium configuration containing a $q$-soliton. The energy of this
configuration is given by:%
\begin{eqnarray*}
E\left[ \left\{ \sigma _{0},\varphi _{0}\right\} \right] &=&E_{\text{\textsc{%
m}}}\left[ \left\{ \sigma _{0},\varphi _{0}\right\} \right] +E_{\text{%
\textsc{i}}}\left[ \left\{ \sigma _{0},\varphi _{0}\right\} \right] +E_{%
\text{\textsc{f}}}\left[ \left\{ \sigma _{0},\varphi _{0}\right\} \right] \\
&=&E\left[ \sigma _{0}\right] +E_{\text{\textsc{i}}}\left[ \left\{ \sigma
_{0},\varphi _{0}\right\} \right] +E_{\text{\textsc{f}}}\left[ \varphi _{0}%
\right]
\end{eqnarray*}%
where 
\begin{eqnarray}
E\left[ \sigma _{0}\right] &=&\int \left[ \frac{1}{2}\left\vert \nabla
u_{0}\right\vert ^{2}-\frac{1}{2}\omega _{0}^{2}u_{0}^{2}dx\right] dx+\int
W(u_{0})dx;\   \label{Es} \\
E_{\text{\textsc{i}}}\left[ \left\{ \sigma _{0},\varphi _{0}\right\} \right]
&=&\omega _{0}\beta \int \varphi _{0}u_{0}dx;  \notag \\
E_{\text{\textsc{f}}}\left[ \varphi _{0}\right] &=&\frac{1}{2}\int
\left\vert \nabla \varphi _{0}\right\vert ^{2}dx.  \notag
\end{eqnarray}%
All these terms are positive; $E\left[ \sigma _{0}\right] $ is positive by (%
\ref{me}) and (\ref{wlo}); $E_{\text{\textsc{i}}}\left[ \left\{ \sigma
_{0},\varphi _{0}\right\} \right] $ is positive since, by eq. (\ref{poisson}%
) $\omega _{0}$ and $\varphi _{0}(x)$ have same sign. Thus this term is
positive also for anti-solitons. The energy $E\left[ \sigma _{0}\right] +E_{%
\text{\textsc{i}}}\left[ \left\{ \sigma _{0},\varphi _{0}\right\} \right] $
is concentrated around $0$ in a region of radius $r_{0}.$ In fact, since $u$
decays exponentially, from the physical point of view, it can be considered
null for $\left\vert x\right\vert $ larger that a suitable $r_{0}.$ The
field energy $E_{\text{\textsc{f}}}\left[ \varphi _{0}\right] $ is not
concentrated; by eq. (\ref{poisson}), it decays as $\left\vert x\right\vert
^{-4}$. Finally, notice that the e.m. field energy of a soliton does not
diverge as the energy of a pointwise particle would do.

\subsection{Travelling $q$-solitons}

The action functional is invariant for the group of the Lorentz boosts:%
\begin{equation}
\ t^{\prime }=\frac{t-vx_{1}}{\sqrt{1-v^{2}}};\ x^{\prime }=\left( \frac{%
x_{1}-vt}{\sqrt{1-v^{2}}},\ x_{2},\ x_{3}\right) .  \label{x1}
\end{equation}%
Hence if $u(t,x),$ $S(t,x),\ \varphi (t,x),$ $\mathbf{A}(t,x)$ is a solution
of (\ref{e1}),..,(\ref{e4}), also $u(t^{\prime },x^{\prime }),$ $S(t^{\prime
},x^{\prime }),\ \varphi ^{\prime }(t^{\prime },x^{\prime }),\mathbf{A}%
^{\prime }(t,x)$ is a solution.

Since $(\varphi ,\mathbf{A})$ is a $4$-vector it transforms as follows:%
\begin{equation*}
\varphi ^{\prime }(t,x)=\frac{\varphi \left( t^{\prime },x^{\prime }\right)
-vA_{1}\left( t^{\prime },x^{\prime }\right) }{\sqrt{1-v^{2}}}
\end{equation*}%
\begin{equation*}
\mathbf{A}^{\prime }(t,x)=\left( \frac{A_{1}\left( t^{\prime },x^{\prime
}\right) -v\varphi \left( t^{\prime },x^{\prime }\right) }{\sqrt{1-v^{2}}},\
A_{2}\left( t^{\prime },x^{\prime }\right) ,\ A_{3}\left( t^{\prime
},x^{\prime }\right) \right)
\end{equation*}%
As usual we set%
\begin{equation*}
\gamma =\frac{1}{\sqrt{1-v^{2}}}
\end{equation*}%
If $\sigma _{0}$ denotes the stationary $q$-soliton defined by (\ref{q-s})
and if $\mathbf{v}=(v,0,0),$ we get the following family of solutions:%
\begin{equation}
u_{\mathbf{v}}(t,x):=u_{0}\left( \gamma \left( x_{1}-vt\right)
,x_{2},x_{3}\right) =u_{0}(x^{\prime })  \label{giove}
\end{equation}%
\begin{equation}
S_{\mathbf{v}}(t,x):=-\omega _{0}\gamma \left( t-vx_{1}\right) =-\omega
_{0}t^{\prime }=\mathbf{k}\cdot x-\omega _{\mathbf{v}}t  \label{stella}
\end{equation}%
where%
\begin{eqnarray}
\mathbf{k} &=&\left( k,0,0\right) =\left( \gamma \omega _{0}v,0,0\right) ;\
\ \   \label{giove+} \\
\omega _{\mathbf{v}} &=&\gamma \omega _{0};  \label{giove++}
\end{eqnarray}%
\begin{equation}
\varphi _{\mathbf{v}}(t,x)=\gamma \varphi _{0}\left( x^{\prime }\right)
\label{marte}
\end{equation}%
\begin{equation}
\mathbf{A}_{\mathbf{v}}(t,x)=-\gamma \left( v\varphi _{0}(x^{\prime }),\ 0,\
0\right)  \label{marte+}
\end{equation}

\bigskip

If $\mathbf{v=}\left( v,0,0\right) ,$ $\left\vert v\right\vert <1,\ $we
define a moving solitons as follows: 
\begin{equation*}
\sigma _{\mathbf{v}}\left( x\right) =\left[ 
\begin{array}{c}
(u_{\mathbf{v}}(0,x),S_{\mathbf{v}}(0,x),0,0) \\ 
(\partial _{t}u_{\mathbf{v}}(0,x),\partial _{t}S_{\mathbf{v}}(0,x),0,0)%
\end{array}%
\right] =\left[ 
\begin{array}{c}
(u_{0}(x^{\prime }),\mathbf{k},\ 0,\ 0) \\ 
(\partial _{t}u_{0}(x^{\prime }),-\gamma \omega _{0},\ 0,\ 0)%
\end{array}%
\right] _{t=0}.
\end{equation*}%
The configuration 
\begin{equation*}
\sigma _{\mathbf{v}}\left( x\right) +\left[ 
\begin{array}{c}
(0,0,\gamma \varphi _{0}\left( x^{\prime }\right) ,-\gamma \left( v\varphi
_{0}(x^{\prime })\right) ) \\ 
(0,0,\gamma \partial _{t}\varphi _{0}\left( x^{\prime }\right) ,-\gamma
v\partial _{t}\varphi _{0}(x^{\prime }))%
\end{array}%
\right] _{t=0}
\end{equation*}%
is initial condition of the solution (\ref{giove}), (\ref{stella}), (\ref%
{marte}), (\ref{marte+}) of eqs.' (\ref{e1}),..,(\ref{e4}).

If $\mathbf{v}$ is any vector with $\left\vert \mathbf{v}\right\vert <1,$ $%
R\in O(3)$ is a rotation such that $R\mathbf{v=}\left( \left\vert \mathbf{v}%
\right\vert ,0,0\right) ,$ we set,%
\begin{equation*}
\sigma _{\mathbf{v}}\left( x\right) =\sigma _{R_{\mathbf{v}}^{-1}\mathbf{v}%
}\left( x\right)
\end{equation*}

\begin{definition}
\label{MS}A moving $q$-soliton with velocity $\mathbf{v}\in \mathbb{R}^{3}$
in the point $\bar{x}\in \mathbb{R}^{3}$ is a function of the form%
\begin{equation*}
\sigma _{\mathbf{v}}\left( x-\bar{x}\right) .
\end{equation*}
\end{definition}

The evolution of a free moving soliton is given by%
\begin{equation}
\sigma _{\mathbf{v}}\left( x-\mathbf{v}t-\bar{x}\right) =\sigma _{R_{\mathbf{%
v}}^{-1}}\left( x-R\left( \mathbf{v}t-\bar{x}\right) \right) .  \label{MS+}
\end{equation}

\subsection{Mechanical properties of $q$-solitons}

First, we will investigate the intrinsic quantities of a moving $q$-soliton.
Since these properties are independent of $R$ and $\bar{x}$ we will just
consider $\sigma _{\mathbf{v}}$ with $\mathbf{v=}\left( v,0,0\right) $ and $%
\bar{x}=0.$

The simplest quantity to describe of a $q$-soliton is the electric charge.
It is defined by (\ref{ccc}) and in this case is 
\begin{equation}
q\left[ \sigma _{0}\right] =\omega _{0}\beta \int u_{0}dx  \label{ecar+}
\end{equation}%
It depends only on the soliton and not on the configuration of the
surrounding field. Moreover it has the following property:

\begin{proposition}
The electric charge of a moving soliton is independent of the motion:%
\begin{equation*}
q\left[ \sigma _{\mathbf{v}}\right] :=q\left[ \sigma _{0}\right]
\end{equation*}
\end{proposition}

\textbf{Proof}: By (\ref{ccc}) and (\ref{giove++}), making a change of
variable $x_{1}=1/\gamma x_{1}^{\prime }+vt,$ we have that 
\begin{eqnarray*}
q\left[ \sigma _{\mathbf{v}}\right] &=&-\beta \int \partial _{t}S_{\mathbf{v}%
}\ dx=\beta \omega _{\mathbf{v}}\int u_{\mathbf{v}}(0,x)\ dx \\
&=&\gamma \omega _{0}\beta \int u_{\mathbf{v}}(x^{\prime })dx=\gamma \omega
_{0}\beta \int u_{\mathbf{v}}(x^{\prime })\frac{1}{\gamma }dx^{\prime } \\
&=&\omega _{0}\beta \int u_{0}dx.
\end{eqnarray*}

$\square $

\bigskip

From now on $q$ will denote the charge of a $q$-soliton. Next, let us
consider the mass:

\begin{definition}
\label{mass}The \textbf{mass} of a moving $q$-soliton is defined by 
\begin{equation*}
m\left[ \sigma _{\mathbf{v}}\right] :=\frac{\mathbf{P}\left[ \sigma _{%
\mathbf{v}}\right] }{\mathbf{v}}
\end{equation*}%
where $\mathbf{P}\left[ \sigma _{\mathbf{v}}\right] $ is the momentum of the
matter field (see Prop. \ref{mome}), namely:%
\begin{equation*}
\mathbf{P}\left[ \sigma _{\mathbf{v}}\right] =\mathbf{P}_{\text{\textsc{m}}}%
\left[ \left\{ \sigma _{\mathbf{v}},\varphi _{\mathbf{v}}\right\} \right]
=\int \left( \partial _{t}u_{\mathbf{v}}\nabla u_{\mathbf{v}}+\partial
_{t}S_{\mathbf{v}}\nabla Su_{\mathbf{v}}^{2}\right) dx
\end{equation*}
\end{definition}

\begin{remark}
Notice that this definition of mass is intrinsic to the equations (\ref{e1}%
),...,(\ref{e4}) and it is independent of any physical interpretation; it
can be interpreted as a "physical" mass whenever $x$ and $t$ are interpreted
as variables of the physical space-time.
\end{remark}

Let us compute explicitly the momentum:

\begin{theorem}
The \textbf{momentum }of a $q$-soliton takes the following form:%
\begin{equation}
\mathbf{P}_{\text{\textsc{m}}}\left[ \sigma _{\mathbf{v}}\right] =\gamma 
\mathbf{v}\left[ \frac{1}{3}\int \left\vert \nabla u_{0}\right\vert
^{2}dx+\omega _{0}^{2}\int u_{0}^{2}dx\right]  \label{P}
\end{equation}
\end{theorem}

\textbf{Proof}: By Prop. \ref{mome}, 
\begin{equation*}
\mathbf{P}\left[ \sigma _{\mathbf{v}}\right] =\int \left( \partial _{t}u_{%
\mathbf{v}}\nabla u_{\mathbf{v}}+\partial _{t}S_{\mathbf{v}}\nabla S\ u_{%
\mathbf{v}}^{2}\right) dx;
\end{equation*}%
assuming $\mathbf{v}=\left( 1,0,0\right) ,$by (\ref{giove}),...,(\ref%
{giove++}) 
\begin{eqnarray*}
\mathbf{P}_{1}\left[ \sigma _{\mathbf{v}}\right] &=&\int \left( \partial
_{t}u_{\mathbf{v}}\partial _{x_{1}}u_{\mathbf{v}}+\partial _{t}S_{\mathbf{v}%
}\partial _{x_{1}}S\ u_{\mathbf{v}}^{2}\right) dx \\
&=&\int \partial _{t}u_{0}(x^{\prime })\partial _{x_{1}}u_{0}(x^{\prime
})dx\ +k\omega _{\mathbf{v}}\int u_{0}^{2}(x^{\prime })dx \\
&=&\int \left[ \partial _{x_{1}^{\prime }}u_{0}(x^{\prime })\partial
_{t}x_{1}^{\prime }\right] \left[ \partial _{x_{1}^{\prime }}u_{0}(x^{\prime
})\partial _{x_{1}}x_{1}^{\prime }\right] dx+v\gamma ^{2}\omega _{0}^{2}\int
u_{0}^{2}(x^{\prime })dx \\
&=&v\gamma ^{2}\int \left[ \partial _{x_{1}^{\prime }}u_{0}(x^{\prime })%
\right] ^{2}dx+v\gamma ^{2}\omega _{0}^{2}\int u_{0}^{2}(x^{\prime })dx
\end{eqnarray*}%
Making a change of variable $x_{1}=1/\gamma x_{1}^{\prime }+vt,$ we get 
\begin{eqnarray*}
\mathbf{P}_{1}\left[ \sigma _{\mathbf{v}}\right] &=&v\gamma ^{2}\int \left[
\partial _{x_{1}^{\prime }}u_{0}(x^{\prime })\right] ^{2}\frac{1}{\gamma }%
dx^{\prime }+v\gamma \omega _{0}^{2}\int u_{0}^{2}(x^{\prime })\frac{1}{%
\gamma }dx^{\prime } \\
&=&v\gamma \left[ \int \left[ \partial _{x_{1}}u_{0}(x)\right] ^{2}dx+\omega
_{0}^{2}\int u_{0}^{2}(x)dx\right]
\end{eqnarray*}%
Since $u_{0}$ is radially symmetric, 
\begin{equation*}
\int \partial _{x_{1}}u_{0}^{2}dx=\frac{1}{3}\int \left\vert \nabla
u_{0}\right\vert ^{2}dx
\end{equation*}%
then 
\begin{equation*}
\mathbf{P}_{1}\left[ \sigma _{\mathbf{v}}\right] =v\gamma \left[ \frac{1}{3}%
\int \left\vert \nabla u_{0}\right\vert ^{2}dx+\omega _{0}^{2}\int
u_{0}^{2}(x)dx\right]
\end{equation*}%
It is immediate to see that $\mathbf{P}_{2}\left[ \sigma _{\mathbf{v}}\right]
=$ $\mathbf{P}_{3}\left[ \sigma _{\mathbf{v}}\right] =0$ and hence we get
the conclusion.

$\square $

So we have obtained the following result:

\begin{corollary}
The \textbf{mass }of a $q$-soliton takes the following value:%
\begin{equation}
m\left[ \sigma _{\mathbf{v}}\right] =\gamma m\left[ \sigma _{0}\right]
=\gamma \left[ \frac{1}{3}\int \left\vert \nabla u_{0}\right\vert
^{2}dx+\omega _{0}^{2}\int u_{0}^{2}dx\right]  \label{massa}
\end{equation}
\end{corollary}

From now on $m$ will denote the rest mass of a $q$-soliton.

We define the energy of a moving soliton as follows:%
\begin{equation*}
E\left[ \sigma _{\mathbf{v}}\right] =E_{\text{\textsc{m}}}\left[ \left\{
\sigma _{\mathbf{v}},\varphi _{\mathbf{v}}\right\} \right]
\end{equation*}

The next proposition describes how the energy transforms in a moving soliton:

\begin{theorem}
\label{marlen}The energy of a $q$-soliton is given by 
\begin{equation*}
E\left[ \sigma _{\mathbf{v}}\right] :=\gamma m+\frac{1}{\gamma }\left( \frac{%
5}{3}\omega _{0}\beta \int \varphi _{0}u_{0}dx\right) =\gamma m-\frac{1}{%
\gamma }\left( \frac{5}{3}\int \varphi _{0}\rho _{0}dx\right)
\end{equation*}%
where $\rho _{0}(x)=\omega _{0}\beta u_{0}(x)$ (see (\ref{ccc})).
\end{theorem}

\begin{remark}
In a theory with $\beta =0,$ the energy of a soliton coincides with its mass 
$\gamma m$ and hence it transforms as the time-component of a time-like
vector. If $\beta \neq 0$ part of the energy transforms differently. This
fact is not so surprising since the energy of a $q$-soliton includes the
energy of the selfinteraction of the soliton with the e.m. field generated
by itself. The energy-momentum of the e.m. field does not transform as the
energy of a space-like vector since it is a light-like vector. Hence there
is a term which is small of the order $\beta $ which transforms differently.
Since this terms is related to the interaction of the matter field with the
e.m. field it might be related to a sort of classical counterpart of the
fine-structure constant; however this point needs a further investigation.
\end{remark}

In order to prove Th. \ref{marlen} we need the following lemma which is a
variant of the Pohozaev-Derrik theorem (\cite{D64},\cite{Poho}):

\begin{lemma}
\label{tderrik}If $u$ is any solution of eqs. (\ref{KG}), (\ref{poisson}),
then%
\begin{equation*}
\int W(u)dx=\frac{1}{2}\omega ^{2}\int u^{2}dx-\frac{1}{6}\int \left\vert
\nabla u\right\vert ^{2}dx-\frac{5}{3}\omega ^{2}\beta ^{2}\int \left( G\ast
u\right) udx
\end{equation*}
\end{lemma}

\textbf{Proof}: Let 
\begin{equation}
J\left[ u\right] =\int \left[ \frac{1}{2}\left\vert \nabla u\right\vert
^{2}+W(u)dx-\frac{1}{2}\omega ^{2}u^{2}dx+\omega ^{2}\beta ^{2}\left( G\ast
u\right) u\right] dx  \label{E8}
\end{equation}%
be the functional $J$ defined by (\ref{J}). Then, if $u$ is a solution of
eqs. (\ref{KG}), (\ref{poisson}), we have that $dJ\left[ u\right] =0.$ Now,
let us consider the "curve" $\lambda \longmapsto u_{\lambda }$ in $V=H^{1}+%
\mathcal{D}^{1.2}$ defined by 
\begin{equation*}
u_{\lambda }=u\left( \frac{x}{\lambda }\right)
\end{equation*}%
Then, 
\begin{equation*}
\left( \frac{d}{d\lambda }J\left[ u_{\lambda }\right] \right) _{\lambda
=1}=0.
\end{equation*}%
Making the change of variable $x\longmapsto x\lambda ^{-1},$ we get that%
\begin{equation*}
J\left[ u_{\lambda }\right] =\frac{\lambda }{2}\int \left\vert \nabla
u\right\vert ^{2}dx+\lambda ^{3}\int W(u)dx-\frac{1}{2}\lambda ^{3}\int
\omega ^{2}u^{2}dx+\lambda ^{5}\beta ^{2}\int \omega ^{2}\left( G\ast
u\right) udx
\end{equation*}%
Then,%
\begin{eqnarray*}
0 &=&\frac{d}{d\lambda }J\left[ u_{\lambda }\right] _{\lambda =1} \\
&=&\left[ \frac{1}{2}\int \left\vert \nabla u\right\vert ^{2}dx+3\lambda
^{3}\int W(u)dx-\frac{3}{2}\lambda ^{2}\int \omega ^{2}u^{2}dx+5\lambda
^{4}\beta ^{2}\int \omega ^{2}\left( G\ast u\right) udx\right] _{\lambda =1}
\\
&=&\frac{1}{2}\int \left\vert \nabla u\right\vert ^{2}dx+3\int W(u)dx-\frac{3%
}{2}\int \omega ^{2}u^{2}dx+5\beta ^{2}\int \omega ^{2}\left( G\ast u\right)
udx
\end{eqnarray*}%
Hence%
\begin{equation*}
\int W(u)dx=\frac{1}{2}\int \omega ^{2}u^{2}dx-\frac{1}{6}\int \left\vert
\nabla u\right\vert ^{2}dx-\frac{5}{3}\omega ^{2}\beta ^{2}\int \left( G\ast
u\right) udx.
\end{equation*}

$\square $

\begin{corollary}
\label{cor}Given a stationary $q$-soliton $\sigma _{0},$ we have that 
\begin{equation*}
E\left[ \sigma _{0}\right] =\frac{1}{3}\int \left\vert \nabla
u_{0}\right\vert ^{2}dx+\omega _{0}^{2}\int u_{0}^{2}dx-\frac{5}{3}\beta
^{2}\int \left( G\ast u_{0}\right) u_{0}dx
\end{equation*}
\end{corollary}

\textbf{Proof}: Replacing $W$ in (\ref{Es}), and using Lemma \ref{tderrik},
we get:%
\begin{eqnarray*}
E\left[ \sigma _{0}\right] &=&\frac{1}{2}\int \left[ \left\vert \nabla
u_{0}\right\vert ^{2}+\omega _{0}^{2}u_{0}^{2}\right] dx+\frac{1}{2}\int
W(u_{0})dx \\
&=&\frac{1}{2}\int \left[ \left\vert \nabla u_{0}\right\vert ^{2}+\omega
_{0}^{2}u_{0}^{2}\right] dx+\frac{1}{2}\int \omega ^{2}u^{2}dx \\
&&-\frac{1}{6}\int \left\vert \nabla u\right\vert ^{2}dx-\frac{5}{3}\omega
^{2}\beta ^{2}\int \left( G\ast u_{\lambda }\right) u_{\lambda }dx \\
&=&\left( \frac{1}{2}-\frac{1}{6}\right) \int \left\vert \nabla
u_{0}\right\vert ^{2}dx+\left( \frac{1}{2}+\frac{1}{2}\right) \omega
_{0}^{2}\int u_{0}^{2}dx-\frac{5}{3}\beta ^{2}\int \left( G\ast u_{0}\right)
u_{0}dx
\end{eqnarray*}

$\square $

\bigskip

\textbf{Proof of Th.\ref{marlen}}: By Prop. \ref{E} and (\ref{giove}),...,(%
\ref{marte+}) we have that 
\begin{eqnarray*}
E\left[ \sigma _{\mathbf{v}}\right] &=&\frac{1}{2}\int \left\vert \partial
_{t}u_{0}(x^{\prime })\right\vert ^{2}dx+\frac{1}{2}\int \left\vert \nabla
u_{0}(x^{\prime })\right\vert ^{2}dx \\
&&+\frac{1}{2}\left( k^{2}+\omega _{\mathbf{v}}^{2}\right) \int
u_{0}(x^{\prime })^{2}dx+\int W\left( u_{0}(x^{\prime })\right) dx
\end{eqnarray*}%
making the change of the integration variable $x_{1}=1/\gamma x_{1}^{\prime
}+vt$, we get%
\begin{eqnarray*}
E\left[ \sigma _{\mathbf{v}}\right] &=&\frac{1}{2\gamma }\int \left\vert
\partial _{t}u_{0}(x^{\prime })\right\vert ^{2}dx^{\prime }+\frac{1}{2\gamma 
}\int \left\vert \nabla u_{0}(x^{\prime })\right\vert ^{2}dx^{\prime } \\
&&+\frac{1}{2\gamma }\left( k^{2}+\omega _{\mathbf{v}}^{2}\right) \int
u_{0}(x^{\prime })^{2}dx^{\prime }+\frac{1}{\gamma }\int W\left(
u_{0}(x^{\prime })\right) dx^{\prime }
\end{eqnarray*}%
Let us compute each piece individually: 
\begin{eqnarray*}
A &=&\frac{1}{2\gamma }\int \left\vert \partial _{t}u_{0}(x^{\prime
})\right\vert ^{2}dx^{\prime }=\frac{1}{2}\int \left\vert \partial
_{x_{1}^{\prime }}u_{0}(x^{\prime })\partial _{t}x_{1}^{\prime }\right\vert
^{2}dx^{\prime } \\
&=&\frac{1}{2}\int \left\vert \partial _{x_{1}^{\prime }}u_{0}(x^{\prime
})\gamma ^{2}v^{2}\right\vert ^{2}dx^{\prime }=\frac{v^{2}\gamma }{2}\int
\left\vert \partial _{x_{1}^{\prime }}u_{0}(x^{\prime })\right\vert
^{2}dx^{\prime } \\
&=&\frac{v^{2}\gamma }{2}\int \left\vert \partial
_{x_{1}}u_{0}(x)\right\vert ^{2}dx
\end{eqnarray*}%
Since $u$ is radially symmetric, 
\begin{equation}
\int \left\vert \partial _{x_{i}}u_{0}(x)\right\vert ^{2}dx=\frac{1}{3}\int
\left\vert \nabla u_{0}\right\vert ^{2}dx  \label{mina}
\end{equation}%
Then,%
\begin{equation*}
A=\frac{v^{2}\gamma }{6}\int \left\vert \nabla _{x^{\prime
}}u_{0}\right\vert ^{2}dx
\end{equation*}%
Let us compute the second piece using (\ref{mina}) again : 
\begin{eqnarray*}
B &=&\frac{1}{2\gamma }\int \left\vert \nabla u_{0}(x^{\prime })\right\vert
^{2}dx^{\prime } \\
&=&\frac{1}{2\gamma }\int \left[ \left\vert \partial _{x_{1}^{\prime
}}u_{0}(x^{\prime })\partial _{x_{1}}x_{1}^{\prime }\right\vert ^{2}\right]
dx^{\prime }+\frac{1}{2\gamma }\int \left[ \left\vert \partial
_{x_{2}^{\prime }}u_{0}(x^{\prime })\right\vert ^{2}+\left\vert \partial
_{x_{3}^{\prime }}u_{0}(x^{\prime })\right\vert ^{2}\right] dx^{\prime } \\
&=&\frac{1}{2\gamma }\int \left\vert \partial _{x_{1}^{\prime
}}u_{0}(x^{\prime })\gamma \right\vert ^{2}dx^{\prime }+\frac{1}{2\gamma }%
\int \left[ \left\vert \partial _{x_{2}^{\prime }}u_{0}(x^{\prime
})\right\vert ^{2}+\left\vert \partial _{x_{3}^{\prime }}u_{0}(x^{\prime
})\right\vert ^{2}\right] dx^{\prime } \\
&=&\frac{\gamma }{6}\int \left\vert \partial _{x_{1}}u_{0}(x)\right\vert
^{2}dx+\frac{1}{2\gamma }\int \left[ \left\vert \partial
_{x_{2}}u_{0}(x)\right\vert ^{2}+\left\vert \partial
_{x_{3}}u_{0}(x)\right\vert ^{2}\right] dx \\
&=&\frac{\gamma }{6}\int \left\vert \nabla u_{0}\right\vert ^{2}dx+\frac{1}{%
3\gamma }\int \left\vert \nabla u_{0}\right\vert ^{2}dx=\left( \frac{\gamma 
}{6}+\frac{1}{3\gamma }\right) \int \left\vert \nabla u_{0}\right\vert ^{2}dx
\end{eqnarray*}%
In order to compute the third piece, we need (\ref{giove+}) and (\ref%
{giove++}): 
\begin{eqnarray*}
C &=&\frac{1}{2\gamma }\left( k^{2}+\omega _{\mathbf{v}}^{2}\right) \int
u_{0}(x)^{2}dx=\frac{1}{2\gamma }\left[ \left( \gamma \omega _{0}v\right)
^{2}+\left( \gamma \omega _{0}\right) ^{2}\right] \int u_{0}(x)^{2}dx \\
&=&\frac{1}{2}\omega _{0}^{2}\gamma \left( v^{2}+1\right) \int u_{0}(x)^{2}dx
\end{eqnarray*}%
The computation of the fourth piece uses Lemma \ref{tderrik}:

\begin{eqnarray*}
\frac{1}{\gamma }\int W\left( u_{0}(x^{\prime })\right) dx^{\prime } &=&%
\frac{1}{\gamma }\int W\left( u_{0}(x)\right) dx \\
&=&-\frac{1}{6\gamma }\int \left\vert \nabla u_{0}\right\vert ^{2}dx+\frac{1%
}{2\gamma }\omega _{0}^{2}\int u_{0}^{2}dx-\frac{5}{3}\omega _{0}^{2}\beta
^{2}\int \left( G\ast u_{0}\right) u_{0}dx \\
&=&E+F+G
\end{eqnarray*}%
Then, 
\begin{equation*}
E\left[ \sigma _{\mathbf{v}}\right] =A+B+C+E+F=\left( A+B+E\right) +\left(
C+F\right) +G
\end{equation*}%
We have that%
\begin{equation*}
\gamma ^{2}v^{2}+\gamma ^{2}+1=\frac{v^{2}+1}{1-v^{2}}+1=\frac{2}{1-v^{2}}%
=2\gamma ^{2}
\end{equation*}%
then,%
\begin{eqnarray*}
A+B+E &=&\left( \frac{v^{2}\gamma }{6}+\frac{\gamma }{6}+\frac{1}{3\gamma }-%
\frac{1}{6\gamma }\right) \int \left\vert \nabla u_{0}\right\vert ^{2}dx \\
&=&\frac{1}{6\gamma }\left( v^{2}\gamma ^{2}+\gamma ^{2}+1\right) \int
\left\vert \nabla u_{0}\right\vert ^{2}dx \\
&=&\frac{\gamma }{3}\int \left\vert \nabla u_{0}\right\vert ^{2}dx
\end{eqnarray*}%
and%
\begin{eqnarray*}
C+F &=&\left[ \frac{1}{2}\omega _{0}^{2}\gamma \left( v^{2}+1\right) +\frac{1%
}{2\gamma }\omega _{0}^{2}\right] \int u_{0}(x)^{2}dx \\
&=&\frac{1}{2}\frac{\omega _{0}^{2}}{\gamma }\left[ \gamma ^{2}\left(
v^{2}+1\right) +1\right] \int u_{0}(x)^{2}dx \\
&=&\frac{1}{2}\frac{\omega _{0}^{2}}{\gamma }2\gamma ^{2}\int
u_{0}(x)^{2}dx=\gamma \omega _{0}^{2}\int u_{0}(x)^{2}dx
\end{eqnarray*}%
Concluding, using Cor. \ref{cor}, we have that%
\begin{equation*}
E\left[ \sigma _{\mathbf{v}}\right] =\frac{\gamma }{3}\int \left\vert \nabla
u_{0}\right\vert ^{2}dx+\gamma \omega _{0}^{2}\int u_{0}(x)^{2}dx-\frac{5}{%
3\gamma }\omega _{0}^{2}\beta ^{2}\int \left( G\ast u_{0}\right) u_{0}dx
\end{equation*}

$\square $

\bigskip

Placing a stationary $q$-soliton in an generic electromagnetic field with
gauge potential $\left( \varphi ,\mathbf{A}\right) $ and using the notation (%
\ref{q-s}), we get the following configuration,%
\begin{equation*}
\sigma _{0}+\left[ 
\begin{array}{c}
\left( 0,0,\varphi ,\mathbf{A}\right) \\ 
\left( 0,0,\partial _{t}\varphi ,\partial _{t}\mathbf{A}\right)%
\end{array}%
\right] =\left[ 
\begin{array}{c}
\left( u_{0},0,\varphi ,\mathbf{A}\right) \\ 
\left( 0,-\omega _{0},\partial _{t}\varphi ,\partial _{t}\mathbf{A}\right)%
\end{array}%
\right] ;
\end{equation*}%
By Prop. \ref{E}, the energy of this configuration is%
\begin{eqnarray*}
E\left[ \left\{ \sigma _{0},\varphi ,\mathbf{A},\partial _{t}\mathbf{A}%
\right\} \right] &=&E_{\text{\textsc{m}}}\left[ \left\{ \sigma _{0},\varphi
\right\} \right] +E_{\text{\textsc{i}}}\left[ \left\{ \sigma _{0},\varphi ,%
\mathbf{A},\partial _{t}\mathbf{A}\right\} \right] +E_{\text{\textsc{f}}}%
\left[ \left\{ \varphi ,\mathbf{A},\partial _{t}\mathbf{A}\right\} \right] \\
&=&E\left[ \sigma _{0}\right] +\omega _{0}\beta \int \varphi \ u_{0}dx+\frac{%
1}{2}\int \left( \mathbf{E}^{2}+\mathbf{H}^{2}\right) dx
\end{eqnarray*}%
If the soliton is small with respect to $\nabla \varphi ,$ (namely if $%
r_{0}=k\ell $ is small), then, by (\ref{ecar+}),%
\begin{equation*}
\omega _{0}\int \varphi (x)u_{0}(x)\ dx\cong \omega _{0}\varphi (0)\beta
\int u_{0}(x)\ dx=q\varphi (0)
\end{equation*}%
where "$\cong $" means that the accuracy of this approximation is good if
the quantities involved are large with respect to $\beta \ $(and to the
radius of the soliton). In fact, the field $\varphi _{0}(x)$ produced by the 
$q$-soliton, is of the order of $\beta \ll 1,$ and hence, if $\varphi
\approx 1,$ we have that $\varphi -\varphi _{0}\cong \varphi $ and $\nabla
\left( \varphi -\varphi _{0}\right) \cong \nabla \varphi .$ Then 
\begin{equation*}
E\left[ \left\{ \sigma _{0},\varphi ,\mathbf{A}\right\} \right] \cong E\left[
\sigma _{0}\right] +q\varphi (0)+\frac{1}{2}\int \left( \mathbf{E}^{2}+%
\mathbf{H}^{2}\right) dx
\end{equation*}%
Therefore, thanks to Prop. \ref{E} and our analysis if a soliton is placed
in a e.m. field we can distinguish the \textbf{soliton energy} $E\left[
\sigma _{0}\right] $, the \textbf{potential energy} $q\varphi (0)$ and the 
\textbf{e.m. field energy} $\frac{1}{2}\int \left( \mathbf{E}^{2}+\mathbf{H}%
^{2}\right) dx$. This distinction is crucial for the study of the dynamics
of the soliton (see section \ref{DS}). Finally, we remark that, the
potential energy $q\varphi (0)$ is localized within the radius of the
soliton. This fact eliminates one of the difficulties posed by the dualism
particle-field where the localization of the potential energy of a particle
is a meaningless problem.

If the $q$-soliton is moving, extending the above arguments, we have the
following result:

\begin{proposition}
\label{pap}If the $q$-soliton is small with respect to $\nabla \varphi $ and 
$\nabla \mathbf{A}$ and $\beta \ll 1$, then 
\begin{eqnarray}
E\left[ \left\{ \sigma _{\mathbf{v}},\varphi ,\mathbf{A}\right\} \right]
&=&E_{\text{\textsc{m}}}\left[ \left\{ \sigma _{\mathbf{v}},\varphi _{%
\mathbf{v}}\right\} \right] +E_{\text{\textsc{f}}}\left[ \left\{ \sigma _{%
\mathbf{v}},\varphi _{\mathbf{v}},\mathbf{A}\right\} \right]  \notag \\
&\cong &\gamma m+q\left[ \varphi (0)+\mathbf{v\cdot A}(0)\right]
\label{pap+}
\end{eqnarray}
\end{proposition}

\textbf{Proof}: By Th. \ref{marlen}, $E_{\text{\textsc{m}}}\left[ \left\{
\sigma _{\mathbf{v}},\varphi _{\mathbf{v}}\right\} \right] =\gamma E\left[
\sigma _{0}\right] \cong \gamma m.$ Then by Prop. \ref{E} and (\ref{giove+}%
), we get%
\begin{eqnarray*}
E\left[ \left\{ \sigma _{\mathbf{v}},\varphi ,\mathbf{A}\right\} \right]
&=&\gamma m-\int \left( \varphi \partial _{t}S_{\mathbf{v}}+\mathbf{A}\cdot
\nabla S_{\mathbf{v}}\right) u_{\mathbf{v}}dx \\
&=&\gamma m+\omega _{0}\int \varphi (x)u_{\mathbf{v}}\left( x\right) \gamma
dx+\omega _{0}\mathbf{v}\cdot \int \mathbf{A}(x)u_{\mathbf{v}}(x)\gamma dx
\end{eqnarray*}%
and, using (\ref{x1}), (\ref{giove}), (\ref{marte}), (\ref{marte+}) and
making a change of variables, we have that%
\begin{eqnarray*}
E\left[ \left\{ \sigma _{\mathbf{v}},\varphi ,\mathbf{A}\right\} \right]
&=&\gamma m+\omega _{0}\int \varphi (x)u_{0}\left( x^{\prime }\right)
dx^{\prime }+\omega _{0}\mathbf{v}\cdot \int \mathbf{A}(x)u_{0}(x^{\prime
})dx^{\prime } \\
&=&\gamma m+\omega _{0}\int \varphi (L^{-1}x)u_{0}\left( x\right) dx+\omega
_{0}\mathbf{v}\cdot \int \mathbf{A}(L^{-1}x)u_{0}(x)dx
\end{eqnarray*}%
where $L$ denotes the Lorentz boost defined by (\ref{x1}), namely $%
Lx=x^{\prime }.$ If the soliton is small with respect to $\nabla \varphi $
and $\nabla \mathbf{A},$ then, using the definition (\ref{ecar+}) of $q$, 
\begin{equation*}
\omega _{0}\int \varphi (L^{-1}x)u_{0}\left( x\right) dx\cong \varphi
(L^{-1}0)\omega _{0}\int u_{0}\left( x\right) dx=q\varphi (0)
\end{equation*}%
and similarly 
\begin{equation*}
\omega _{0}\mathbf{v}\cdot \int \mathbf{A}(L^{-1}x)u_{0}(x)dx\cong \mathbf{v}%
\cdot \mathbf{A}(L^{-1}0)\omega _{0}\int u_{0}\left( x\right) \ dx=q\varphi 
\mathbf{v}\cdot \mathbf{A}(0)
\end{equation*}

$\square $

\bigskip

Notice that (\ref{pap+}) is the energy of the soliton, namely the matter
field energy plus the interaction energy contained in the radius of the
soliton; the total energy of a configuration which contains a soliton,
depends also on $\partial _{t}\mathbf{A}$ and, by Prop. \ref{E} and Prop \ref%
{pap}, it takes the following form:%
\begin{equation*}
E_{tot}\left[ \left\{ \sigma _{\mathbf{v}},\varphi ,\mathbf{A},\partial _{t}%
\mathbf{A}\right\} \right] \cong \gamma m+q\left[ \varphi (0)+\mathbf{v\cdot
A}(0)\right] +\frac{1}{2}\int \left( \mathbf{E}^{2}+\mathbf{H}^{2}\right) dx.
\end{equation*}

Now let us examine a configuration containing several solitons 
\begin{equation*}
\sigma _{\mathbf{v}_{k},\bar{x}_{k}}:=\sigma _{\mathbf{v}}(\cdot -\bar{x}%
_{k}),\ \ k=1,..,N
\end{equation*}
where $\sigma _{\mathbf{v}_{k}}(\cdot -\bar{x}_{k})$ has been defined by
Def. \ref{MS}. We assume that%
\begin{equation}
\left\vert \bar{x}_{k}-\bar{x}_{h}\right\vert \geq 2r_{0},\ \ k\neq h
\label{far}
\end{equation}%
where $r_{0}$ denote the radius of the solitons. We remember that $u$ decays
exponentially, so the matter field is essentially null out of a neighborhood
of each soliton and hence%
\begin{equation}
E\left[ \sum_{k=1}^{N}\sigma _{\mathbf{v}_{k},\bar{x}_{k}}\right] \cong
\sum_{k=1}^{N}E\left[ \sigma _{\mathbf{v}_{k},\bar{x}_{k}}\right] \cong
m\sum_{k=1}^{N}\gamma _{k}  \label{ms}
\end{equation}%
where%
\begin{equation*}
\gamma _{k}=\frac{1}{\sqrt{1-\left\vert \mathbf{v}_{k}\right\vert ^{2}}}
\end{equation*}%
Notice that, in the configuration (\ref{ms}), also the $q$-antisolitons can
be included. They have the same mass of solitons, but opposite electric
charge.

If we embed this configuration in an external e.m. field, the total energy
takes the following form: 
\begin{eqnarray*}
&&E_{tot}\left[ \left\{ \sum_{k=1}^{N}\sigma _{\mathbf{v}_{k},\bar{x}%
_{k}},\varphi ,\mathbf{A},\partial _{t}\mathbf{A}\right\} \right] \\
&\cong &m\sum_{k=1}^{N}\gamma _{k}+q\sum_{k=1}^{N}\left[ \varphi (\bar{x}%
_{k})+\mathbf{v}_{k}\mathbf{\cdot A}(\bar{x}_{k})\right] +\frac{1}{2}\int
\left( \mathbf{E}^{2}+\mathbf{H}^{2}\right) dx.
\end{eqnarray*}

\subsection{Dynamics of $q$-solitons\label{DS}}

Now let examine the dynamics of a solitons in the presence of an "external"
electromagnetic field. More exactly, we want to examine the behavior of the
solution of the Cauchy problem with the following initial conditions:%
\begin{equation}
U_{0}=\sum_{k=1}^{N}\sigma _{\mathbf{v}_{k},\bar{x}_{k}}+\left[ 
\begin{array}{c}
\left( 0,0,\varphi _{0},\mathbf{A}_{0}\right) \\ 
\left( 0,0,\varphi _{1},\mathbf{A}_{1}\right)%
\end{array}%
\right]  \label{ic}
\end{equation}%
where $\mathbf{v}_{k}\in \mathbb{R}^{3}$ is such that $\left\vert \mathbf{v}%
_{k}\right\vert <1.$

It is well known that, thanks to the invariance of the hylenic ratio, the
soliton is orbitally stable (see e.g. \cite{befolibro}). This means that if
the perturbation field generated by $(\varphi _{0}(x),\mathbf{A}_{0}(x)),$ $%
(\varphi _{1}(x),\mathbf{A}_{1}(x))$ is small (with respect to $\beta ^{-1})$
around the soliton, then the solution of the Cauchy problem has the
following form:%
\begin{equation}
U_{0}(t,x)=\sum_{k=1}^{N}\sigma _{\mathbf{v}_{k}(t),\bar{x}_{k}(t)}+\left[ 
\begin{array}{c}
\left( u_{p}(t,x),\ S_{p}(t,x),\varphi (t,x),\mathbf{A}(t,x)\right) \\ 
\left( \partial _{t}u_{p}(t,x),\ \partial _{t}S_{p}(t,x),\partial
_{t}\varphi (t,x),\partial _{t}\mathbf{A}(t,x)\right)%
\end{array}%
\right]  \label{C}
\end{equation}%
where

\begin{itemize}
\item $u_{p}(t,x),\ S_{p}(t,x)$ are essentially null thanks to the orbital
stability of the soliton and they will be neglected;

\item $\sum_{k=1}^{N}\sigma _{\mathbf{v}_{k}(t),\bar{x}_{k}(t)}$ is the
configuration of the $q$-solitons and its structure is determined by a $N$
function $\xi _{k}:\mathbb{R}\rightarrow \mathbb{R}^{3}$ such that $\xi
_{k}\left( t\right) =\bar{x}_{k}(t);\ \ \dot{\xi}_{k}\left( t\right) =%
\mathbf{v}_{k}\left( t\right) ;$
\end{itemize}

Our aim to investigate the dynamics of the $q$-solitons under the following
assumptions:

\begin{itemize}
\item \textit{(A-1)} $\beta \ll 1$; as we have seen this condition implies
that the Cauchy problem is well posed and that the energy of a $q$-solitons
equals its mass (Th. \ref{marlen});

\item \textit{(A-2) }the solitons are far from each other (i.e. (\ref{far})
holds) during the time interval considered; this happens if

\begin{itemize}
\item \textit{(i)} this assumption is satisfied by the initial condition (%
\ref{ic});

\item \textit{(ii) }all the $q$-solitons have the same charge (namely there
are not $q$-antisolitons), so that, during the evolution, the $q$-solitons
repel each other;

\item \textit{(iii) }the e.m. field is not locally too strong, so that the $%
q $-solitons cannot collide;
\end{itemize}

\item \textit{(A-3)} $\left\vert \ddot{\xi}_{k}\left( t\right) \right\vert
\ll 1;$ this fact avoids the $q$-soliton to produce a strong radiation and,
from the technical point of view, it simplify the computations. Clearly this
happens if the e.m. field is not too strong
\end{itemize}

\bigskip

We will show, that under these assumptions the $q$-solitons behave as
classical particles. To this aim, we analyze the action functional relative
to the configuration (\ref{C}) : 
\begin{eqnarray}
\mathbb{A} &=&\iint \left( \mathcal{L}_{\text{\textsc{m}}}+\mathcal{L}_{%
\text{\textsc{i}}}+\mathcal{L}_{\text{\textsc{f}}}\right) dxdt
\label{bella+} \\
&=&\iint \mathcal{L}_{\text{\textsc{m}}}\left[ \sum_{k=1}^{N}\sigma _{\dot{%
\xi}_{k}\left( t\right) ,\xi _{k}\left( t\right) }\right] dxdt  \notag \\
&&+\iint \mathcal{L}_{\text{\textsc{i}}}\left[ \left\{ \sum_{k=1}^{N}\sigma
_{\dot{\xi}_{k}\left( t\right) ,\xi _{k}\left( t\right) },\varphi ,\mathbf{A}%
\right\} \right] dxdt+\iint \mathcal{L}_{\text{\textsc{f}}}\left[ \left\{
\varphi ,\mathbf{A}\right\} \right] dxdt  \notag
\end{eqnarray}%
Since we have assumed \textit{(A-2)}, then%
\begin{equation*}
\int \mathcal{L}_{\text{\textsc{m}}}dx\cong \sum_{k=1}^{N}\int \mathcal{L}_{%
\text{\textsc{i}}}\left[ \sigma _{\dot{\xi}_{k}\left( t\right) ,\xi
_{k}\left( t\right) };\varphi ,\mathbf{A}\right] dx
\end{equation*}%
and%
\begin{equation*}
\int \mathcal{L}_{\text{\textsc{i}}}dx\cong \sum_{k=1}^{N}\int \mathcal{L}_{%
\text{\textsc{i}}}\left[ \left\{ \sigma _{\dot{\xi}_{k}\left( t\right) ,\xi
_{k}\left( t\right) },\varphi ,\mathbf{A}\right\} \right] dx
\end{equation*}

Let us compute each piece of the action separately:

\begin{lemma}
Under the assumptions \textit{(A-1),(A-2),(A-3), }we have that%
\begin{equation*}
\int \mathcal{L}_{\text{\textsc{m}}}\left[ \sigma _{\dot{\xi}_{k}\left(
t\right) ,\xi _{k}\left( t\right) }\right] dx\cong -m\sqrt{1-\left\vert \dot{%
\xi}_{k}\left( t\right) \right\vert ^{2}}.
\end{equation*}
\end{lemma}

\textbf{Proof}: Since the Lagrangian $\mathcal{L}_{\text{\textsc{m}}}$ does
not depend explicitly on $t$ and $x,$ we can choose a reference frame where,
for a fixed $t$, 
\begin{equation*}
\xi _{k}\left( t\right) =0\ \ \ and\ \ \ \dot{\xi}_{k}\left( t\right)
=(v_{k},0,0)
\end{equation*}%
so that 
\begin{equation*}
\int \mathcal{L}_{\text{\textsc{m}}}\left[ \sigma _{\dot{\xi}_{k}\left(
t\right) ,\xi _{k}\left( t\right) }\right] dx=\int \mathcal{L}_{\text{%
\textsc{m}}}\left[ \sigma _{v_{k}}\left( x\right) \right] dx
\end{equation*}%
We recall that by (\ref{giove}) and (\ref{stella}),%
\begin{equation*}
\sigma _{v_{k}}\left( x\right) =\left[ 
\begin{array}{c}
(u_{0}(x^{\prime }),-\gamma _{k}\omega _{0},0,0) \\ 
(0,\gamma _{k}\omega _{0}v_{k},0,0)%
\end{array}%
\right]
\end{equation*}%
where we have set%
\begin{equation*}
\gamma _{k}=\frac{1}{\sqrt{1-\left\vert \dot{\xi}_{k}\left( t\right)
\right\vert ^{2}}}.
\end{equation*}%
Then, by (\ref{2}), and (\ref{giove}),...,(\ref{giove++}) 
\begin{eqnarray*}
\int \mathcal{L}_{\text{\textsc{m}}}\left[ \sigma _{\dot{\xi}_{k}\left(
t\right) ,\xi _{k}\left( t\right) }\right] dx &=&\int \mathcal{L}_{\text{%
\textsc{m}}}\left[ \sigma _{v_{k}}\right] dx \\
&=&\frac{1}{2}\int \left\vert \partial _{t}u_{0}(x^{\prime })\right\vert
^{2}dx-\int \left\vert \nabla u_{0}(x^{\prime })\right\vert ^{2}dx \\
&&+\frac{1}{2}\left( k_{k}^{2}-\omega _{\mathbf{v}_{k}}^{2}\right) \int
u_{0}(x^{\prime })^{2}dx-\int W\left( u_{0}(x^{\prime })\right) dxdt
\end{eqnarray*}%
If we assume that $\ddot{\xi}_{k}$ is not too large (i.e. \textit{(A-2)}),%
\begin{equation*}
\partial _{t}x_{1}^{\prime }=\partial _{t}\frac{x^{\prime }-\dot{\xi}%
_{k}\left( t\right) t}{\sqrt{1-\left\vert \dot{\xi}_{k}\left( t\right)
\right\vert ^{2}}}\cong v_{k}\gamma _{k};
\end{equation*}%
then, arguing as in the proof of Th. \ref{marlen} and using similar
notations for each $k$,%
\begin{equation*}
\int \mathcal{L}_{\text{\textsc{m}}}\left[ \sigma _{\dot{\xi}_{k}\left(
t\right) ,\xi _{k}\left( t\right) }\right]
dx=A_{k}-B_{k}+C_{k}^{a}-C_{k}^{b}-E_{k}-F_{k}-G_{k}
\end{equation*}%
where 
\begin{equation*}
C_{k}^{a}=\frac{1}{2}\omega _{0}^{2}\gamma _{k}v_{k}^{2}\int
u_{0}(x)^{2}dx;\ \ \ C_{k}^{b}=\frac{1}{2}\omega _{0}^{2}\gamma _{k}\int
u_{0}(x)^{2}dx.
\end{equation*}%
Going on with our computation, 
\begin{eqnarray*}
A_{k}-B_{k}-E_{k} &=&\left[ \frac{\gamma _{k}v_{k}^{2}}{6}-\frac{\gamma _{k}%
}{6}-\frac{1}{3\gamma _{k}}+\frac{1}{6\gamma _{k}}\right] \int \left\vert
\nabla u_{0}\right\vert ^{2}dx \\
&=&\frac{1}{6\gamma _{k}}\left[ v_{k}^{2}\gamma _{k}^{2}-\gamma _{k}^{2}-1%
\right] \int \left\vert \nabla u_{0}\right\vert ^{2}dx \\
&=&\frac{1}{6\gamma _{k}}\left[ \frac{v_{k}^{2}-1}{1-v_{k}^{2}}-1\right]
\int \left\vert \nabla u_{0}\right\vert ^{2}dx=-\frac{1}{3\gamma _{k}}\int
\left\vert \nabla u_{0}\right\vert ^{2}dx
\end{eqnarray*}%
\begin{eqnarray*}
C_{k}^{a}-C_{k}^{b}-F_{k} &=&\left[ \frac{1}{2}\omega _{0}^{2}\gamma
_{k}\left( v_{k}^{2}-1\right) -\frac{1}{2\gamma _{k}}\omega _{0}^{2}\right]
\int u_{0}^{2}dx \\
&=&\left[ -\frac{1}{2\gamma _{k}}\omega _{0}^{2}-\frac{1}{2\gamma _{k}}%
\right] \int u_{0}(x)^{2}dx \\
&=&-\frac{\omega _{0}^{2}}{\gamma _{k}}\int u_{0}(x)^{2}dx
\end{eqnarray*}%
The term $G_{k}$ will be ignored since we have assumed $\beta \ll 1$ (i.e. 
\textit{(A-2}). Then, by (\ref{massa}),%
\begin{eqnarray*}
\int \mathcal{L}_{\text{\textsc{m}}}\left[ \sigma _{\dot{\xi}_{k}\left(
t\right) ,\xi _{k}\left( t\right) }\right] dx &=&-\frac{1}{3\gamma _{k}}\int
\left\vert \nabla u_{0}\right\vert ^{2}dx-\frac{\omega _{0}^{2}}{\gamma _{k}}%
\int u_{0}(x)^{2}dx \\
&=&-\frac{1}{\gamma _{k}}m=-m\sqrt{1-\left\vert \dot{\xi}_{k}\left( t\right)
\right\vert ^{2}}
\end{eqnarray*}

$\square $

\bigskip

Now let us compute $\int \mathcal{L}_{\text{\textsc{i}}}dx$.

\begin{lemma}
If \textit{(A-1) and (A-2) hold, then }%
\begin{equation*}
\int \mathcal{L}_{\text{\textsc{I}}}\left[ \sigma _{\dot{\xi}_{k}\left(
t\right) ,\xi _{k}\left( t\right) };\varphi ,\mathbf{A}\right] dx\cong q%
\left[ \varphi \left( t,\xi _{k}(t)\right) -\mathbf{A}\left( t,\xi
_{k}(t)\right) \cdot \dot{\xi}_{k}(t)\right] dt.
\end{equation*}%
\bigskip
\end{lemma}

\textbf{Proof}: As in the previous lemma, we can choose a reference frame
where, for a fixed $t$, $\dot{\xi}_{k}\left( t\right) =\mathbf{v}$ and $\xi
_{k}\left( t\right) =0$. Then following the same arguments used in the proof
of Prop. \ref{pap}, we have that%
\begin{equation*}
\int \mathcal{L}_{\text{\textsc{i}}}\left[ \sigma _{\mathbf{v},\bar{x}%
};\varphi ,\mathbf{A}\right] dx\cong q\left( \mathbf{v\cdot A}(\bar{x}%
)-\varphi (\bar{x})\right) .
\end{equation*}

$\square $

\bigskip

The above lemmas give the following result:

\begin{theorem}
\label{figo}Let $U_{0}(t,x)$ be the solution of the Cauchy problem relative
to equation (\ref{e1}),...,(\ref{e4}) with the initial condition (\ref{ic}).
Then If \textit{(A-1), (A-2) and (A-3) hold, we have that }%
\begin{equation}
\frac{d}{dt}\left( \frac{m\dot{\xi}_{k}}{\sqrt{1-\left\vert \dot{\xi}%
_{k}\right\vert ^{2}}}\right) \cong q\left( \mathbf{E}+\dot{\xi}_{k}\times 
\mathbf{H}\right)  \label{L}
\end{equation}%
\begin{equation*}
\nabla \cdot \mathbf{E}=\sum_{k=1}^{N}\rho _{0}(x-\xi _{k})
\end{equation*}%
\begin{equation*}
\nabla \times \mathbf{H}-\partial _{t}\mathbf{E}=\sum_{k=1}^{N}\mathbf{j}%
_{0}(x-\xi _{k})
\end{equation*}%
\begin{equation*}
\nabla \times \mathbf{E}+\partial _{t}\mathbf{H}=0
\end{equation*}%
\begin{equation*}
\nabla \cdot \mathbf{H}=0
\end{equation*}
\end{theorem}

\textbf{Proof}: The action (\ref{bella+}) becomes 
\begin{eqnarray*}
\mathbb{A} &=&\mathbb{A}_{\text{\textsc{m}}}+\mathbb{A}_{\text{\textsc{i}}}+%
\mathbb{A}_{\text{\textsc{f}}} \\
&=&\sum_{k=1}^{N}-m\int \left[ \sqrt{1-\left\vert \dot{\xi}_{k}\left(
t\right) \right\vert ^{2}}+q_{k}\left[ \mathbf{A}\left( t,\xi _{k}(t)\right)
\cdot \dot{\xi}_{k}(t)-\varphi \left( t,\xi _{k}(t)\right) \right] \right] dt
\\
&&+\iint \mathcal{L}_{\text{\textsc{f}}}\left[ \varphi ,\mathbf{A}\right]
dxdt
\end{eqnarray*}%
Making the variation of $\mathbb{A}_{\text{\textsc{m}}}+\mathbb{A}_{\text{%
\textsc{i}}}$ with respect to $\xi _{k}$, we get the Lorentz equation (\ref%
{L}); making the variation of $\mathbb{A}$ given by (\ref{bella}) we get the
Maxwell equations.

$\square $

\begin{remark}
Theorem \ref{figo} states that equations (\ref{e1}),...,(\ref{e4}) provide a
model for material particles which, at low energies, agree with the well
known physics. It is interesting to investigate the predictions of this
model when the assumptions\textit{\ (A-2), (A-3) are violated. }If
(A-2)-(ii) is violated there are antisolitons which attract solitons since
they have opposite charges; then also (A-2)-(i) will be eventually violated
and the two particles will annihilate. Since our equation is invariant for
time-reversal, also the creation of a couple particle-antiparticle might
occur; of course this can happen only if there is sufficient energy, namely
if (A-2),(iii) does not hold. \textit{If (A-3) does not hold, a numerical
computation of the radiation when }$\ddot{\xi}$ is large gives a spectrum
which probably can be compared with the experimental data.
\end{remark}

\section{Conclusive remarks}

\bigskip

More than 50 years ago, De Broglie wrote:

\begin{quotation}
Des consid\`{e}rations sur lesquelles je reviendrai me conduisent
aujourd'hui a penser que le corpuscule doit etre assimil\'{e} non pas \`{a}
un v\'{e}ritable singularit\'{e} punctuelle de $u$, mais \`{a} un tr\'{e}s
petite r\'{e}gion singuli\'{e}re de l'espace o\`{u} $u$ prenderait une tr%
\`{e}s grande valeur et ob\'{e}irait \`{a} une \'{e}quation non lin\'{e}aire
dont l'\'{e}quation lin\'{e}aire de la M\'{e}canique ondulatoire ne serait
qu'une forme approximative valable en dehors de la region singuli\`{e}re.
L'id\'{e}e que l'\'{e}quation de propagation de $u$, contrairement \`{e} l'%
\'{e}quation classique du $\Psi $, est en principe non lin\'{e}aire
m'apparait meme maintenant comme tout \`{a} fait essentielle. (\cite{DB},
Chap. IX, 1, p. 95.)
\end{quotation}

The development of the nonlinear analysis of the last half century allows to
construct models of particles in line with the ideas of De Broglie. The
model presented here is strongly based on Classical Mechanics and "\textit{a
priori}" has nothing to do with Quantum Mechanics (QM), in contrast with the
ideas of De Broglie. Nevertheless, it is interesting to notice that it
presents some feature which are considered peculiar of QM.

The first thing to remark is the fact that particles-like solutions of
nonlinear equations with positive energy, in dimension 3, seems possible
only if they have at least one internal degree of freedom, namely $\psi $
takes values in $\mathbb{C}$ and not in $\mathbb{R}$ (see e.g Derrik theorem 
\cite{D64}). This fact implies that%
\begin{equation*}
\psi (t,x)=u(t,x)e^{i\left( \mathbf{k\cdot }x-\omega t\right) }
\end{equation*}%
presents an undulatory aspect as desired by De Broglie. Furthermore, since
the energy/momentum $(E,\mathbf{p})$ of the particle and the wave number $%
\left( \omega ,\mathbf{k}\right) $ are $4$-vectors they must be proportional
and hence 
\begin{equation*}
E=\hslash \omega \ \ \ and\ \ \ \mathbf{p}=\hslash \mathbf{k}
\end{equation*}%
where $\hbar $ is a constant depending on the parameters of the problem. So
we can say that eqs'(\ref{e1}),...,(\ref{e4}) present one kind of intrinsic
Plank constant. However, this similarity does not imply the De Broglie pilot
wave theory or the Bohmian mechanics since the "interference" or the
"entanglement" phenomena cannot be reproduced by this model.

The second remarkable fact is the existence of anti-particles, and the fact
that an antiparticle is produced by time-reversion (or by charge inversion).

An other peculiarity is that the $q$-solitons are equal to each other and
two of them cannot be in the same position. This fact implies that they are
forced to follow the Bose-Einstein statistics which also is considered a
quantum phenomenon.

However, we do not think that the $q$-solitons could be considered as a
model for elementary particles. They are just an example (and probably the
simplest one) that shows the possibility of a classical theory of
electrodynamics and the fact that some quantum phenomena are consequences of
a consistent field theory idependently of the quantization.

Nevertheless it is possible to implement the ideas presented here to build a
"classical" model of elementary particles. It is necessary to take $\psi $%
-functions with spinor values and a Lagrangian with a suitable symmetry. For
example in \cite{B2020} a $U(1)\times SU(2)$ symmetry is considered.

The final conclusion is the following: if the Maxwell equations are weakly
coupled in the simplest way with a linear equation, invariant for the Poincar%
\'{e} group, then a small nonlinear perturbation (see Th. \ref{lilli}) is
sufficient to produce non only a consistent electrodynamics theory, but also
solitons which share some characteristic with quantum particles.

\end{document}